\documentclass[]{interact}

\newif\ifblind
\blindfalse

\newcommand{\revised}[1]{\ifblind\textcolor{red}{#1}\else#1\fi}

\usepackage{xcolor}
\usepackage{color}
\usepackage{float}
\usepackage{booktabs}
\usepackage{soul}
\usepackage{amsmath}
\usepackage{bbm}
\usepackage{graphicx}
\usepackage{amssymb}
\usepackage{graphicx,subfig}
\usepackage{epstopdf}
\usepackage{setspace}
\usepackage[linesnumbered,ruled]{algorithm2e}

\usepackage[unicode=true,
 bookmarks=false,
 breaklinks=false,pdfborder={0 0 1},backref=false,colorlinks=true]
 {hyperref}
\hypersetup{
 breaklinks,hyperindex,citecolor=blue,linkcolor=blue}

\makeatletter

\floatstyle{ruled}
\newfloat{algorithm}{tbp}{loa}
\providecommand{\algorithmname}{Algorithm}
\floatname{algorithm}{\protect\algorithmname}

\@ifundefined{date}{}{\date{}}

\usepackage{pdfcomment}
\usepackage{comment}

\usepackage{environ}
\RenewEnviron{comment}{\pdfcomment{\BODY}}

\usepackage{soul}

\usepackage{amsthm}\usepackage{multirow}
\usepackage{multicol}\usepackage{mathrsfs}
\usepackage{caption}

\usepackage{tikz}
\usetikzlibrary{fit,positioning,arrows,automata,calc}
\usetikzlibrary{external}
\usepackage{pgfplots}

\usepackage{lipsum}

\graphicspath{{Figures/}}

 \graphicspath{{figures/}} 

\usepackage{epstopdf}
\usepackage{natbib}
\bibpunct[, ]{(}{)}{;}{a}{}{,}

\begin{document}
\title{Partially Observable Markov Decision Process Framework for Operating Condition Optimization Using Real-Time Degradation Signals}
    
\ifblind
\else
\author{
\name{Boyang Xu, Yunyi Kang, Xinyu Zhao, Hao Yan, Feng Ju\textsuperscript{a}}
\affil{\textsuperscript{a}Arizona State University, Tempe, AZ, 85281, the United States}
}
\thanks{E-mail of the authors are boyangxu@asu.edu, ykang37@asu.edu, xzhao119@asu.edu, HaoYan@asu.edu, fengju@asu.edu.}
\fi  

\markboth{JQT}{}
\maketitle

\maketitle

\begin{abstract}

In many engineering systems, proper predictive maintenance and operational control are essential to increase efficiency and reliability while reducing maintenance costs. However, one of the major challenges is that many sensors are used for system monitoring. Analyzing these sensors simultaneously for better predictive maintenance optimization is often very challenging. In this paper, we propose a systematic decision-making framework to improve the system performance in manufacturing practice, considering the real-time degradation signals generated by multiple sensors. Specifically, we propose a partially observed Markov decision process (POMDP) model to generate the optimal capacity and predictive maintenance policies, given the fact that the observation of the system state is imperfect. Such work provides a systematic approach that focuses on jointly controlling the operating conditions and preventive maintenance utilizing the real-time machine deterioration signals by incorporating the degradation constraint and non-observable states. We apply this technique to the bearing degradation data and NASA aircraft turbofan engine dataset, demonstrating the effectiveness of the proposed method.
\end{abstract}




\section{Introduction}
\label{sec:introduction}
Manufacturing systems are often highly dynamic and coupled with many operations
(e.g., machines, robots, and material handling devices), which are subject to status degradation.
In such complex systems, the degradation of components is one of the main reasons for unplanned
downtime, which leads to substantial costs from both system repair and production loss
\cite{shen2017reliability,sun2020managing,kang2019performance}.
In response, preventive maintenance (PM), which emphasizes effectively avoiding the occurrence of
unexpected severe failures, becomes a preferred maintenance choice for practitioners. Typically, a
conditional-based preventive maintenance strategy \cite{butler2017introduction,kang2019flexible}
is often used in the literature \cite{caesarendra2011combined,tobon2012data}. Researchers have
designed sophisticated condition-based maintenance policies that rely on various types of condition
information, providing insights on how to reduce maintenance costs and increase equipment
reliability. The purpose of PM is to manage component degradation so that the
degradation of components and the system is controlled to be below a safety margin.

In addition to preventative maintenance, the degradation process of a unit is directly influenced by
different operating conditions \cite{kang2019joint}. Controllable factors such as the level of usage,
speed, force, load, and operating time have been examined to tie closely to the degradation process
of a machine. For example, in a battery pack, a lighter workload usually results in a slower increase
in the temperature in the battery cells, resulting in a longer time in good use conditions and less
frequent (preventive) maintenance. On the other hand, the lighter workload can also result in less
capacity, which triggers long-term production performance and can make existing replacement
policies cost-ineffective. The wind turbine gearboxes and generators deteriorate faster at higher
speeds, conveyor belts fail more often when used at higher rotational speeds and cutting tools wear
faster at higher speeds. Furthermore, the optimal conditions should also be determined by the
current machine states, which can be inferred from the real-time degradation signals.

Despite extensive research on degradation modeling and preventive maintenance,
important gaps remain. Most remaining useful life (RUL) prediction methods either assume fully
observable degradation or reduce the problem to discrete latent states using standard HMMs,
typically relying on single-sensor signals and lacking explicit physical constraints. Existing
maintenance optimization studies often design policies under simplified dynamics
(e.g., exponential lifetimes) or treat operating conditions as fixed, and preventive maintenance
and capacity planning are usually addressed separately. These limitations hinder adaptability in
realistic multi-sensor, multi-condition environments.

Moreover, although multi-sensor monitoring has been widely studied,
many fusion methods either collapse signals into a single health index or apply independent HMMs
to each sensor, losing cross-sensor correlation and suffering from state misalignment.
Factorial or mixture HMMs can handle multiple channels but often ignore physical monotonic
degradation constraints, leading to unstable training or non-interpretable reverse transitions.

In conclusion, the following challenges exist in the current research: 1) The signals from the sensors are usually imperfect due to the sensing noise. 2) The observations or signals are often  multi-channel with a complicated correlation structure. 3) The optimal maintenance decision is related to the degradation states, which are typically unknown. In fact, according to a survey, on an average of 28 \% of the total production cost is attributed to the mishandling of  operational and maintenance activities in the modern industry \cite{wang2012overview}, which calls out the necessity of efficient studies. To address these challenges, ideally, an intelligent decision-making framework should be formulated, which jointly considers all the operation conditions, integrate the multi-channel degradation signals, dynamically control the workload and maintenance tactics and ultimately maximize the production performance. 


In this paper, we propose a new decision--making framework that schedules operational activities by jointly considering internal machine conditions and external operating settings.
The framework is formulated as a \emph{partially observable Markov decision process (POMDP)} that integrates multi-sensor degradation signals to achieve long--term performance improvement through coordinated capacity control and preventive maintenance.
At its core, the model embeds a \emph{constrained Input-Output Hidden Markov Model (IOHMM)} to capture the machine degradation process under multiple sensors.
The constrained IOHMM enforces a left-to-right physical degradation structure and shares emission distributions across operating conditions, ensuring physically consistent state progression and stable training even in noisy, high-dimensional sensor environments.
This latent-state estimator provides reliable belief updates for the POMDP layer, which then optimizes maintenance and capacity decisions under partial observability.
Traditional HMMs are effective for state estimation but only passively infer degradation stages and cannot directly link state estimation to operational decision-making.
Standard MDPs, on the other hand, assume fully observable states, which is unrealistic in noisy industrial settings.
By combining constrained IOHMM state estimation with POMDP-based control, the proposed architecture bridges the gap between \emph{estimation} and \emph{decision optimization}, offering (i) physically consistent state progression, (ii) robustness to multi-sensor noise, (iii) capacity-dependent degradation captured through distinct transition matrices, and (iv) seamless integration of state inference with optimal preventive maintenance and operating-condition control.
Semi-supervised learning further enables the framework to leverage non--run-to-failure data and exploit information from intermediate states, while a shared set of hidden states allows consistent evaluation of degradation progression across varying operating conditions.

The literature either focuses on the RUL prediction given the degradation signals or provides the preventive maintenance strategy with an observable system state. This paper, on the other hand, provides a systematic approach that focuses on jointly controlling the operating conditions and preventative maintenance  utilizing the real-time machine deterioration signals by incorporating the degradation constraint and non-observable states. 

The main contributions of this work are threefold:

\textbf{1. Integrated Multi-Sensor Degradation Modeling:} We develop a constrained Input-Output Hidden Markov Model (IOHMM) that integrates heterogeneous sensor data while enforcing physically meaningful left-to-right degradation constraints. Unlike existing approaches that either rely on single-sensor signals or ignore physical constraints, our framework captures the complex correlation structure across multiple sensors and ensures realistic monotonic degradation paths.

\textbf{2. Noisy Observation Handling:} We address the challenge of noisy sensor observations through a POMDP-based framework that explicitly models uncertainty in state estimation. Unlike existing approaches that assume perfect observability or rely on simple threshold-based methods, our framework provides robust state inference under noisy conditions, enabling reliable decision-making even when sensor signals are corrupted or incomplete.

\textbf{3. Sequential Real-Time Decision Making:} We develop a complete methodology for sequential real-time control that bridges the gap between theoretical modeling and practical implementation. The framework provides parameter estimation algorithms, real-time state inference, and sequential decision-making procedures that enable adaptive control actions based on evolving system conditions. This addresses the need for dynamic control strategies that can respond to changing system states in real-time manufacturing environments.

We will organize the rest of the paper as follows: Section II reviews the relevant literature. Section III develops the proposed framework and efficient computational algorithms. Section IV demonstrates the effectiveness of the proposed methodology using two case studies: (1) the bearing degradation dataset collected from the XJTU-SY experimental platform \cite{wang2018hybrid}, and (2) the NASA Turbofan Engine Degradation Simulation Data Set (C-MAPSS) \cite{saxena2008turbofan}. Section V discusses the sensitivity analysis and comparison study. Finally, Section VI provides the conclusion and discussion of future research directions.

\section{Literature Review}
\label{sec:literature_review}
In this literature review, we will focus on two major components. 1) Data-driven degradation modeling for RUL estimation. 2) Preventative maintenance and capacity planning strategy. 

In the first group of degradation modeling for RUL estimation, we would like to classify the research into a single operating condition and multiple operating conditions. For a single operating condition, many research works have focused on modeling the degradation process as a continuous process, such as the state-space model \cite{sun2014prognostics}, degradation path model \cite{lu1993using}, and health-index-based approach \cite{yan2016multiple}. The models mentioned assume that the degradation process can be monitored and represented explicitly. 
On the other hand, there are many research represent the degradation states as discrete states due to the abrupt changes between states. More specifically, the Hidden Markov Model (HMM) is a type of discrete state transition model that assumes the Markovian property for the transitions. In literature, inspired by the introductory work by \cite{rabiner2002tutorial}, there is much existing research on applying HMM for accurate fault prediction and RUL estimation in the area of prognostics \cite{soualhi2016hidden,boutros2011detection,liu2014zero}. Hidden Markov Model (HMM) models are effective in capturing the dynamics of a system by utilizing a finite number of latent states. These states are trained based on ordered and timed observations, allowing HMM models to provide insightful explanations for processes with complex state transitions. In many cases, HMM models have a simple structure, often comprising a single layer. \cite{athanasopoulou2010maximum,geramifard2013multimodal}. Some variants of HMM, such as the two-layer model \cite{camci2010health}, factorial hidden Markov model \cite{le2016competing}, hidden semi-Markov model \cite{liu2015novel}, etc have also been proposed. However, these models either work on one single sensor or do not consider the physical left-to-right constraints. Furthermore, there is no guidance on how to make a predictive maintenance policy from these systems.

While traditional HMMs have proven effective for single-condition scenarios, the modeling of degradation under multiple operating conditions requires more sophisticated approaches. Input-Output Hidden Markov Models (IOHMMs) extend traditional HMMs by incorporating external inputs that influence state transitions and emissions \cite{bengio1994input,bengio1996input}. The foundational work introduced IOHMMs as a framework for modeling sequential data where the dynamics depend on external control signals.

For degradation modeling under multiple operating conditions, IOHMMs offer several advantages over traditional HMMs. The work by \cite{gonzalez2005modeling} demonstrated the effectiveness of IOHMMs for analyzing
and forecasting electricity spot prices. Similarly, \cite{rocher2021iohmm} propose drift detection of IoT-Based Systems. However, existing IOHMM applications in degradation modeling typically focus on either single-sensor scenarios or lack explicit physical constraints. Most existing works either only considers single sensing variable. do not enforce the left-to-right constraint that is physically meaningful for degradation processes, potentially leading to unrealistic state transitions that violate the monotonic nature of degradation.

The works by \cite{gebraeel2008prognostic,bian2015degradation} developed a sensor-based degradation model based on the general path model under two operational conditions. The work by \cite{whitmore1997modelling} considered a Wiener diffusion process with a time-scale transformation to address the challenge of multiple operational conditions.

The second group of research focuses on improving system performance and discusses various methodologies for generating maintenance policies. These methodologies are evaluated using different performance benchmarks, such as average operating time \cite{klutke2002availability}, machine cycle time \cite{liao2006maintenance}, operating costs \cite{liyanage2003towards}, and energy consumption \cite{luo2015data}, which aims to maintain a system in certain conditions while it is still operational \cite{alaswad2017review,wang2018condition}, is a common topic within different maintenance practices. Recently, many models are developed to supervise the decision-making procedures, such as determining proper maintenance  intervals, optimizing the consumption of the spare parts, and minimizing related operating costs \cite{wang2012overview}. For example, the work by \cite{douer1994optimal} aims to optimize the PM actions for a single machine, while \cite{keizer2017condition} focuses on the system with multiple machines and complex structures, taking into account variations in machine conditions. For a more in-depth review of the conditional and predictive maintenance, please refer to \cite{jardine2006review} and  \cite{wang2002survey}.

Finally, there are also works related to the operating condition and capacity adjustment using real-time degradation signals. For the work considering the degradation process with multiple operating conditions, paper \cite{yang2007maintenance} jointly considers maintenance scheduling and adjustable throughput in their model. However, the operating time is assumed to follow the exponential distribution, which is not true in many cases.
The works by \cite{hao2015controlling}  and \cite{li2017study} consider the real-time degradation signals when making decisions for capacity adjustment. Nevertheless, the maintenance is simply performed after the machine failure, and no preventative maintenance options are discussed. For the POMDP framework, the work by \cite{byon2010season} develops a condition-based maintenance model to generate the most cost-effective policy  that can be adapted to wind farm operations, considering the seasonal impact on turbine capacities. However, their model treats the alternating capacities as non-controllable parameters and automatically switches throughout the year, which limits its application scenarios. The work by \cite{aldurgam2013optimal} considers the multiple capacity problems, but they assume that all the parameters in the POMDP models are typically provided. They provide no guidelines on utilizing the real-time degradation signals in the framework, which leads to practical concerns. \cite{zhao2019semi} proposed to utilize a Hidden Markov Model in estimating the states of the systems, whereas this paper proposed to optimize the operational conditions in dynamic multi-sensor environments. Recently, \cite{deep2023partially} proposed a POMDP-based optimal maintenance planning model that accounts for time-dependent condition monitoring signals and derives a control-limit policy for preventive maintenance. However, their approach does not consider operating condition adjustments, assumes a single-source degradation signal, and lacks real-time adaptability to dynamic system changes. In contrast, our framework jointly optimizes maintenance and operating decisions, integrates multi-sensor degradation signals, and employs a constrained IOHMM with a left-to-right degradation constraint, enabling more accurate state estimation and adaptive decision-making.

Although extensive research has been conducted on both degradation modeling and preventive maintenance strategies, several important gaps remain. First, most RUL estimation approaches either assume that degradation can be fully observed or reduce the problem to discrete latent states via HMMs. While effective for fault prediction, these models generally rely on single-sensor signals, lack explicit physical degradation constraints, and do not provide a direct bridge to maintenance decision-making. As a result, they are limited in capturing the complexity of real-world systems that operate under varying conditions and rely on multiple heterogeneous sensors. Second, many maintenance optimization approaches have focused on designing policies under simplified assumptions about system dynamics. For example, operating times are often assumed to follow exponential distributions, or capacity changes are treated as uncontrollable parameters. Furthermore, in most cases, maintenance and operational control are addressed separately. Preventive maintenance models tend to prescribe fixed intervals or thresholds, while capacity planning models rarely incorporate real-time degradation signals, limiting their adaptability in dynamic environments. These limitations highlight the need for an integrated framework that can jointly handle degradation modeling and decision optimization under uncertainty.

\section{Methodology}
\label{sec:methodology}
This section presents our comprehensive approach to operational control optimization under partial observability. Our methodology consists of three main components: (1) formulating the operational control problem as a Partially Observable Markov Decision Process (POMDP) in Section \ref{subsec:pomdp_formulation}, which includes system assumptions (Section \ref{subsub:system_assumptions}) and POMDP components definition (Section \ref{subsub:pomdp_components}); (2) modeling the unknown degradation dynamics using a constrained Input-Output Hidden Markov Model (IOHMM) and estimating its parameters from sensor data in Section \ref{subsec:iohmm_estimation}, which encompasses IOHMM architecture (Section \ref{subsub:IOHMM_arch}), parameter estimation via GEM algorithm (Section \ref{subsub:GEM}), and discreting POMDP via GMM in (Section \ref{subsub:discrete}); and (3) solving the POMDP to generate an optimal control policy and implementing it in real-time in Section \ref{subsec:pomdp_solution}, which covers offline policy generation (Section \ref{subsub:offline_policy}) and online real-time control (Section \ref{subsub:online_control}). This structured approach enables us to systematically address the challenges of partial observability, unknown system dynamics, and real-time decision-making in manufacturing systems. 

\subsection{Problem Formulation: A POMDP Framework for Operational Control \label{subsec:pomdp_formulation}}
This subsection formulates the operational control problem as a Partially Observable Markov Decision Process (POMDP). The objective is to find an optimal policy that maps real-time observations to actions (changing operating conditions or performing maintenance) to maximize long-term rewards. We first establish the system assumptions and then formally define the POMDP components.

\subsubsection{System Assumptions \label{subsub:system_assumptions}}
In this paper, we build up a discrete model to depict the operation and degradation process of the machine. The time epochs are denoted as $n = 1,2,\ldots, T$. These time intervals within two-time epochs are set as the machine cycle time, which is assumed to be the time that is required to finish processing the assigned units of product. In our study, all decisions may include capacity adjustment or preventive maintenance that are made and executed only at the beginning of each epoch. Below, we summarize the definitions and assumptions of the studied system. 

\begin{figure}
    \centering
    \includegraphics[width=0.8\linewidth]{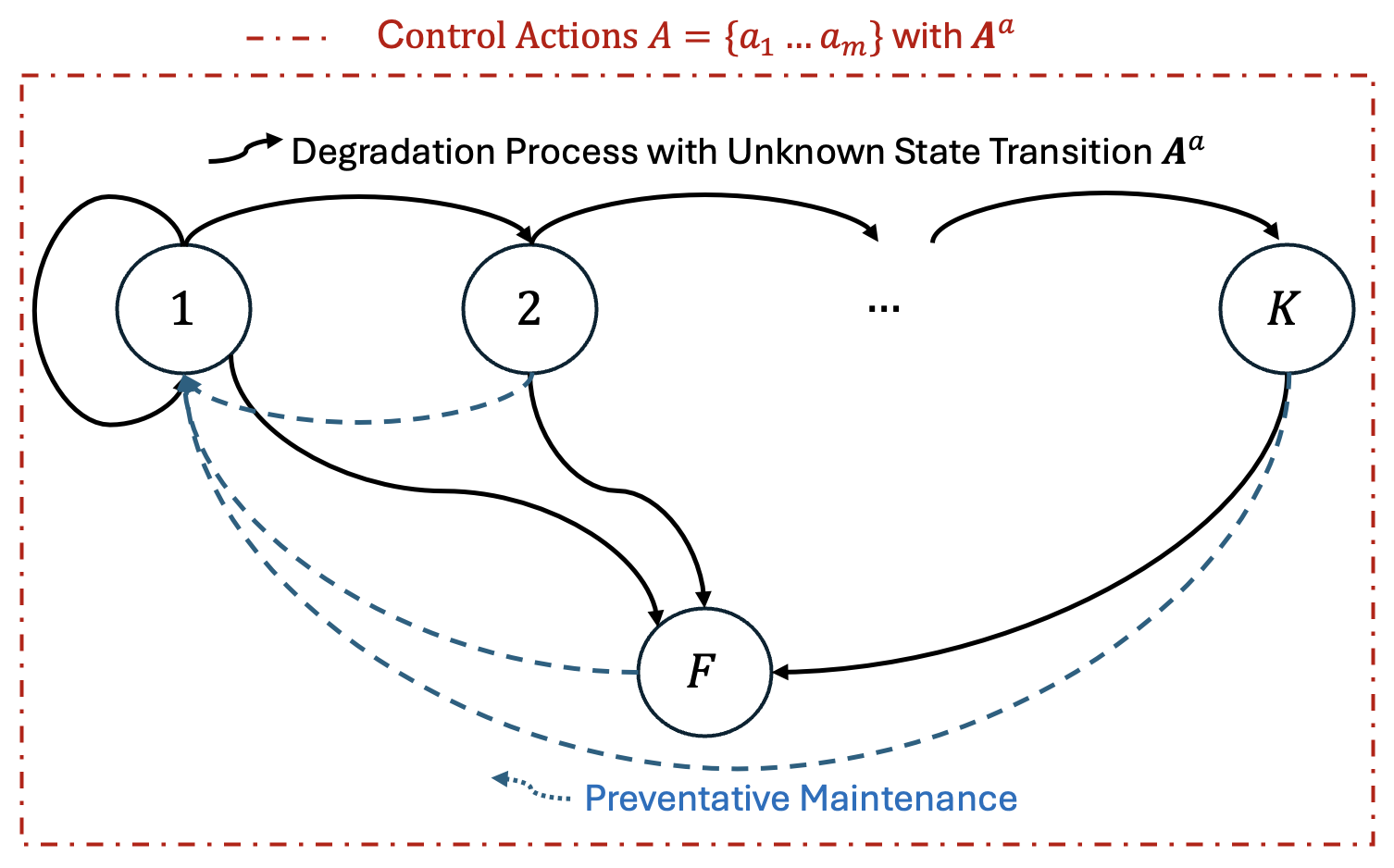}
    \caption{The original system state transitions}
    \label{fig:Original_System_State}
\end{figure}

\begin{itemize}
    \item Machine degradation process: we assume the state $\mathbf{S}=\{1,2, \ldots, K,F\}$ with $K$ operating states and one failure state (denoted by $F$). The transition matrix $X$ between the states is assumed to be static and unknown. For the operating states, they fall into the increasing order that state $1$ represents the best operating state ("brand new") and state $K$ is the worst operating state before the system fails. Furthermore, state $F$ denotes the failure state, which is the last and non-operating state in the degradation process. The transitions of machine states in the degradation process are assumed to occur at the beginning of a time epoch. We assume that the condition of the machine is retaining or decreasing from time to time. For the machine at an operating state $k$ at time $t$, it can retain at state $k$ or transfer to any inferior state at time $t+1$.  In other words, no condition reverse is allowed in the machine degradation process. The machine degradation process is expressed with the solid lines as shown in Figure \ref{fig:Original_System_State}. It is worth noting that we use discrete state representation of the degradation process compared to the continuous state representation. One main reason is that the discrete state representation is good enough to model the non-continuous degradation process, which appears in the study of many systems \cite{gu2016real}. Furthermore, from the control perspective, it is much easier to derive an  interpretative and easy-to-implement policy from the discrete state representation. 
    \item Control actions 1: Maintenance decision. When the machine is failed, the operation must be stopped, and maintenance must be performed, replacing the failed component and recovering the machine conditions thoroughly (usually known as "corrective maintenance"). We assume that such replacement and repair can be finished within one production cycle, and the machine condition will be recovered to the best state (state $1$). Apart from passive corrective maintenance, the decision-maker has the option to choose to conduct preventative maintenance (PM), when the machine is in any of the operating states ($1,\ldots,{K}$) at the beginning of a time slot. If PM is selected, then the machine is stopped immediately and the degrading component in the machine will be replaced. The selection of PM will lead to the early retirement and discard of the component. We also assume that the replacement is conducted within one cycle. The newly installed component is "brand new", which recovers the machine condition to the best, or state $1$. The dashed blue arrow in Figure \ref{fig:Original_System_State} represent these PM transitions. We further assume that preventative maintenance will incur a small cost and machine breakdown will incur a large cost. More definitions will be discussed in Section \ref{subsec:pomdp_formulation}. 
    \item Control actions 2: Capacity adjustment. The machine is also assumed able to operate under different conditions throughout the lifetime, i.e., with different operating capacities, and the operator can switch the capacity when needed at the beginning of a production cycle. Let $\mathbf{A}=\{a_1,\ldots,a_m\}$ denote different operating actions, e.g., the set of all speed options for the machine. We assume that the speed adjustment options are finite, i.e., the different levels on switchers or gears.  Correspondingly, the degradation process under each action varies. However, we assume that the degradation process can be represented using the same set of states $\mathbf{S}$ and may follow different state transition probabilities for different capacities. \revised{Each capacity setting induces its own state-matrix $\mathbf{A}(a)$, as indicated by the red dashed box in ~Figure \ref{fig:Original_System_State}.}
\end{itemize}



\subsubsection{POMDP Components \label{subsub:pomdp_components}}
In practice, the states of the system are non-observable and thus cannot be evaluated directly. Instead, it can be estimated from the signals which monitor the conditions of the machines in real-time. To model and solve the problem, we develop a Partially Observable Markov Decision Process (POMDP) model to solve the optimize the operating condition problem. 

A POMDP is a generalization of a Markov decision process (MDP), which models an agent decision process in which the system dynamics is modeled by an MDP, but the agent cannot directly observe the underlying state space. Here, we choose a discrete-time and discrete-state Markov model to represent the system dynamics. The state-space of the system contains all the states in the machine. In literature, POMDP is a 7-tuple $(S,A,T,R,\Omega,O,\gamma)$, where:

\begin{itemize}
    \item $S$ is a set of states, here we refer to the non-observable health state of the machine.
    \item $A$ is a set of actions. In our context, actions correspond to selecting different operating conditions (e.g., different machine speeds, loads) or performing maintenance decisions (e.g., preventive maintenance, corrective maintenance). Note that in our POMDP formulation, each action $a \in A$ represents a specific operating condition $a \in \mathbf{A}$. 
    \item $T$ is a set of conditional transition probabilities between states, which models the dynamics of the degradation process. Following the IOHMM framework, the transition model $T$ is computed as $\mathbf{A}(a)$ for operating condition $a$, where $\mathbf{A}(a)$ is the input-dependent transition matrix for action $a$. Crucially, these probabilities are unknown and will be estimated in Section \ref{subsec:iohmm_estimation}.
    \item $R: S \times A \to \mathbb{R}$ is the reward function, which refers to the maintenance cost. 
    \item $\Omega$ is a set of observations, which is corresponding to the degradation signals observed by single or multiple sensors. 
    \item $O$ is a set of conditional observation probabilities, which models the data generative process given the observation and state. These probabilities are also unknown and will be estimated in Section \ref{subsec:iohmm_estimation}.
    \item $\gamma \in [0, 1]$ is the discount factor, which models how the reward function diminishes over time. 
\end{itemize}

The performance of the machine denoted as $r(S,a)$, is a condition-based cost metric determined by the machine condition $S$ and the action $a$. When the machine is operating under action $a$ (corresponding to a specific operating condition), the production reward is evaluated by the capacity that the machine provides. When the machine fails, the corrective maintenance cost is occurred, which is denoted as $c_d$. The cost of maintenance $c_p$ is assumed to be uniform for all the states. 

\begin{equation}
\label{reward_pj1}
\begin{split}
r(S, a)=\begin{cases}
    c_a,&  \text{if } S \in \{1,\ldots, K\}, \\
    c_d,&  \text{if } S \in F, \\
    c_p, &  \text{if } S \in PM, \\
0,& \text{otherwise,}
\end{cases}
\end{split}
\end{equation}

Having established the POMDP framework, the next challenge is to learn the unknown transition probabilities $T$ and observation probabilities $O$ from historical sensor data. This is addressed in Section \ref{subsec:iohmm_estimation}, where we develop a constrained IOHMM approach to estimate these parameters while enforcing physical constraints on the degradation process.

\begin{figure}
    \centering
    \includegraphics[width=1\linewidth]{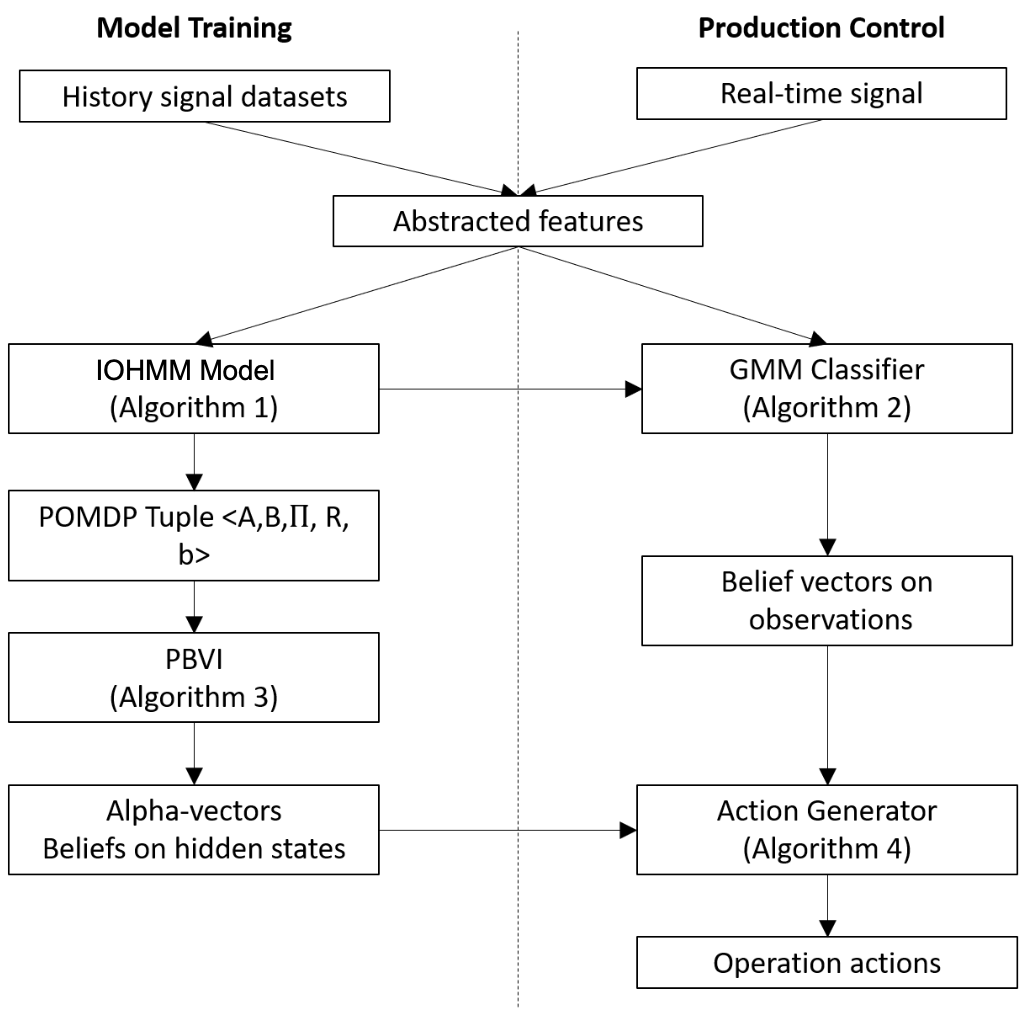}
    \caption{The structure of the algorithms}
    \label{fig:algo_structure}
\end{figure}

The overall solution scheme is summarized as shown in Figure \ref{fig:algo_structure}. For the offline training, as shown on the left side, historical data sets are collected to abstract the features and thereafter used to obtain the parameters for the IOHMM model and the POMDP model, which will ultimately return the vectors for decision-making given different observations. When the model is conducted for real-time control, the signal is abstracted into the features, which will be classified as the belief vectors on the discrete observations. The belief vector, together with the alpha vectors generated from the training stage, will be used to determine the ultimate control action at the exact time.

\subsection{Degradation Modeling and Parameter Estimation via IOHMM \label{subsec:iohmm_estimation}}
This subsection consolidates all aspects of learning the unknown model parameters (transition probabilities $T$ and observation probabilities $O$) from sensor data. We introduce the constrained Input-Output Hidden Markov Model (IOHMM) as the method for modeling the degradation dynamics and learning the unknown POMDP parameters. The "Inputs" are the operating conditions (actions) and the "Outputs" are the sensor observations.

POMDP can be treated as IOHMM with controllable action. In the training phase, we assume that all control actions are given (i.e., the system has recorded which action to use). However, the dynamics under each condition are unknown. In this subsection, we will study how to estimate the parameters using the IOHMM framework with the Generalized Estimation-Maximization (GEM) algorithm. The IOHMM framework trains all parameters jointly, replacing the need for separate HMMs and shared emission parameters. Semi-supervised learning implies that the proposed method can effectively handle the non-run-to-failure data and utilize the information of intermediate states more effectively. The left-to-right constraint is a physical constraint for most degradation systems, which can dramatically stabilize the training for multi-sensor systems. For different operating conditions, the same set of hidden states is used and trained, which enables effective evaluation of the progression of the degradation process provided the switch of operating conditions.


\begin{figure}
    \centering
    \includegraphics[width=0.75\linewidth]{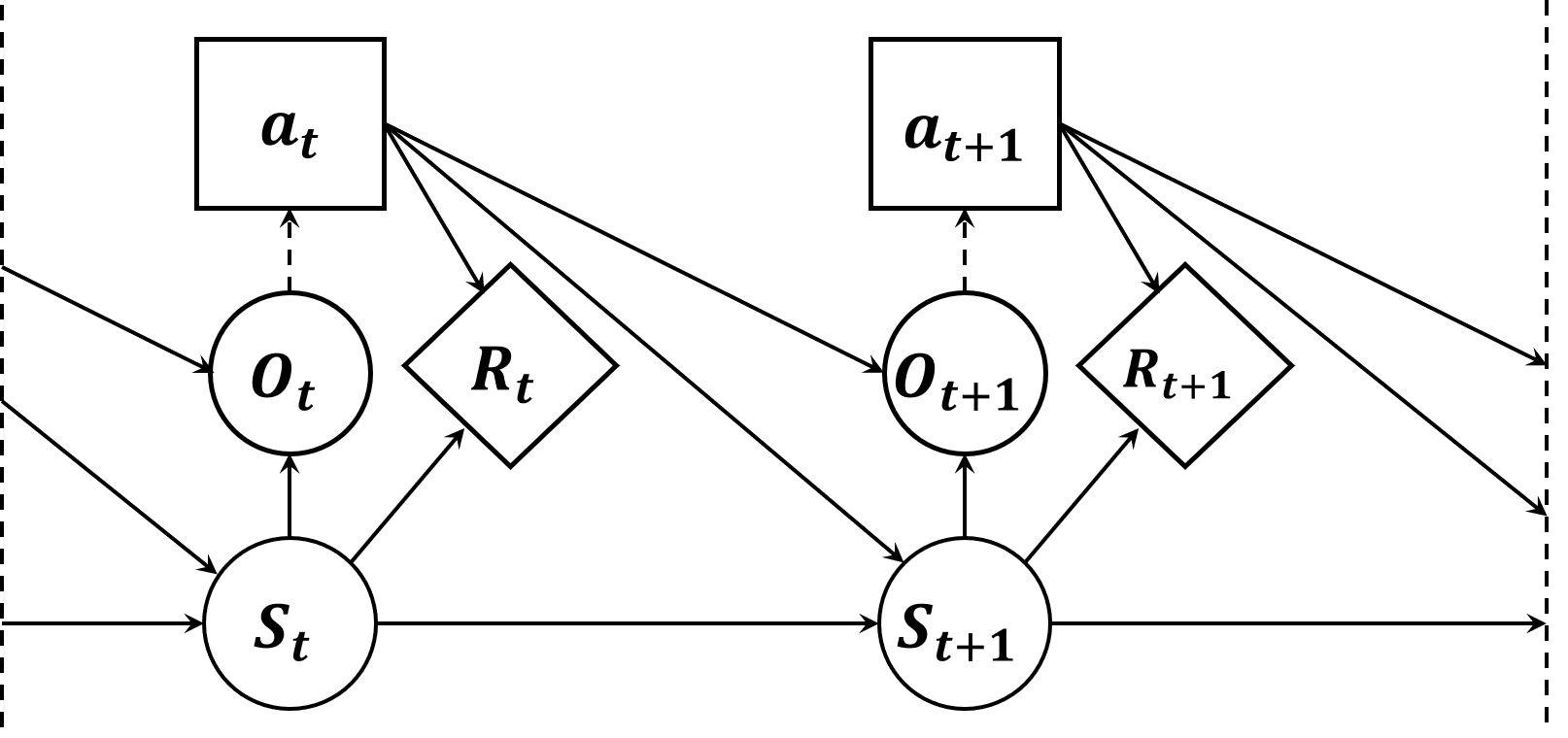}
    \caption{The state transition diagram of POMDP}
    \label{fig:POMDP_demo}
\end{figure}

\subsubsection{Constrained IOHMM Architecture \label{subsub:IOHMM_arch}}
Following the IOHMM framework \cite{bengio1994input,bengio1996input}, the model consists of two main components:

\textbf{1. Input-Dependent Transition Matrices:}
For each action $a \in \mathbf{A}$,
let $\mathbf{A^a} \in \mathbb{R}^{K \times K}$ be the transition matrix whose element
\[
\bigl[\mathbf{A^a}\bigr]_{k\ell}
= P\!\left(S_t = k \,\middle|\, S_{t-1} = \ell,\; A_t = a \right)
\]
gives the probability of moving from state $\ell$ to state $k$ under action $a$.

\textbf{2. Emission Models:} The emission probabilities can be modeled using two different approaches: (1) \textbf{Shared Emission Model}, where emission parameters are common across all actions, represented by $P\left(\mathbf{O}_t \mid S_t=k\right)$; and (2) \textbf{Action-Dependent Emission Model}, where each action has its own emission parameters, represented by $P\left(\mathbf{O}_t \mid S_t=k, A_t=a\right)$.

It is important to note that IOHMM literature distinguishes between two approaches for modeling emissions: (1) \textbf{Shared Emission Matrix}, where emission parameters $\{\boldsymbol{\mu}_k, \boldsymbol{\Sigma}_k\}$ are common across all actions, and (2) \textbf{Action-Dependent Emission Matrix}, where each action $a$ has its own emission parameters $\{\boldsymbol{\mu}_k^a, \boldsymbol{\Sigma}_k^a\}$.

\begin{figure}
    \centering\includegraphics[width=1\linewidth]{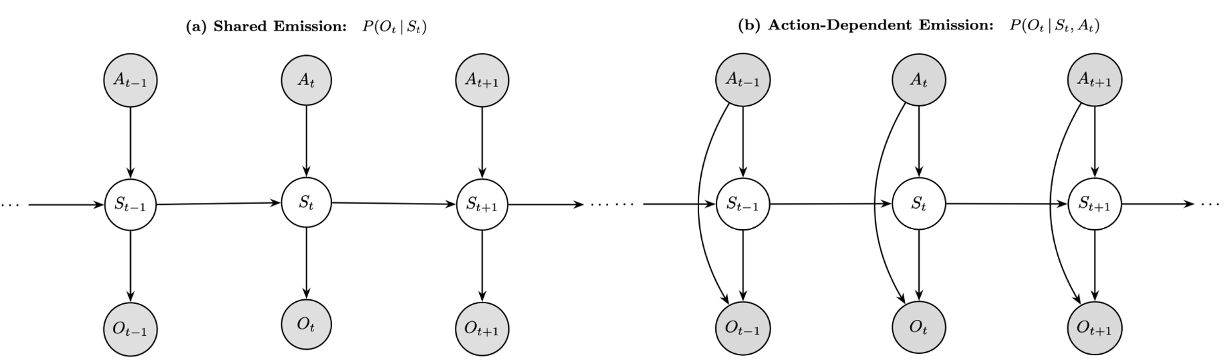}
    \caption{Graphical representation of the IOHMM emission structures.
(a) \emph{Shared emission:} observations follow $P(O_t \mid S_t)$ and
are independent of the action.
(b) \emph{Action--dependent emission:} observations follow
$P(O_t \mid S_t, A_t)$ and thus explicitly depend on the current action.}
    \label{fig:iohmm_frame}
\end{figure}

The shared emission approach assumes that sensor observations under the same hidden state have identical distributions regardless of action, which is appropriate when actions primarily affect transition dynamics rather than emission characteristics. This assumption is valid for our manufacturing application where different machine speeds affect degradation progression rates but the sensor signatures for each degradation level remain consistent.

In contrast, action-dependent emissions are necessary when actions significantly influence sensor readings even for the same underlying degradation state. For example, in aviation applications, engine emissions, vibration patterns, and sensor readings differ substantially for different flight speeds or altitudes even for identical engine health conditions. In such cases, the emission model should be $P(\mathbf{O}_t | S_t=j, A_t=a)$ with action-specific parameters.

Our choice of shared emission parameters is justified by the physical nature of our degradation system, where the fundamental sensor signatures of each degradation state remain consistent across different operating speeds, while only the transition rates between states vary with actions.

The emission probability can be modeled using two different approaches depending on whether actions affect the sensor signatures:

\textbf{Case 1: Shared Emission Matrix (Action-Independent Emissions)}
The emission probability is defined on multi-variate response $\mathbf{O}_{i,t}$ by the shared emission matrix $\mathbf{B} = \{\boldsymbol{\mu}_{k}, \boldsymbol{\Sigma}_{k}\}$, where $\boldsymbol{\mu}_{k}$ is the mean value of multiple sensors in state $k$ (shared across all actions as shown in Figure \ref{fig:iohmm_frame} (a)), and $\boldsymbol{\Sigma}_{k}$ is the covariance matrix of state $k$ (shared across all actions):
    
\begin{equation}
    P\left(\mathbf{O}_{i,t} \mid S_{i,t}=k\right) = \mathcal{N}\left(\mathbf{O}_{i,t}; \boldsymbol{\mu}_{k}, \boldsymbol{\Sigma}_{k}\right)
\end{equation}

\textbf{Case 2: Action-Dependent Emission Matrix (Non-Shared Emissions)}
The emission probability is defined on multi-variate response $\mathbf{O}_{i,t}^a$ by action-specific emission matrices $\mathbf{B}^a = \{\boldsymbol{\mu}_{k}^a, \boldsymbol{\Sigma}_{k}^a\}$, where $\boldsymbol{\mu}_{k}^a$ is the mean value of multiple sensors in state $k$ under action $a$, and $\boldsymbol{\Sigma}_{k}^a$ is the covariance matrix of state $k$ under action $a$ as shown in Figure \ref{fig:iohmm_frame} (b):
    \begin{equation}
    P\left(\mathbf{O}_{i,t}^a \mid S_{i,t}=k, A_{i,t}=a\right) = \mathcal{N}\left(\mathbf{O}_{i,t}^a; \boldsymbol{\mu}_{k}^a, \boldsymbol{\Sigma}_{k}^a\right)
    \end{equation}

The choice between these two approaches depends on the physical characteristics of the system. Case 1 is appropriate when actions primarily affect transition dynamics but not the fundamental sensor signatures of each degradation state. Case 2 is necessary when actions significantly influence sensor readings even for the same underlying degradation state.

Given a dataset $\mathcal{D} = \{(\mathbf{O}_i, \mathbf{A}_i)\}_{i=1}^N$ where $\mathbf{O}_i = \{\mathbf{O}_{i,t}\}_{t=1}^{T_i}$ represents the sequence of observations for unit $i$ and $\mathbf{A}_i = \{A_{i,t}\}_{t=1}^{T_i}$ represents the corresponding sequence of actions, the IOHMM likelihood can be expressed as:

\begin{equation}
\begin{aligned}
& L(\Theta; \mathcal{D}) = P(\mathbf{O} | \mathbf{A}; \Theta) = \prod_{i=1}^{N} P(\mathbf{O}_i | \mathbf{A}_i; \Theta) \\
& P(\mathbf{O}_i | \mathbf{A}_i) = \sum_{S_i} P(\mathbf{O}_i, S_i | \mathbf{A}_i) \\
& = \sum_{S_i} \left( P(S_{i,1} | A_{i,1}) \prod_{t=2}^{T_i} P(S_{i,t} | S_{i,t-1}, A_{i,t}) \prod_{t=1}^{T_i} P(\mathbf{O}_{i,t} | S_{i,t}, A_{i,t}) \right)
\end{aligned}
\end{equation}

where $\Theta$ represents the model parameters (transition and emission parameters), $S_i = \{S_{i,t}\}_{t=1}^{T_i}$ denotes the hidden state sequence for unit $i$, and the likelihood factors into initial state probabilities, state transition probabilities, and emission probabilities, all conditioned on the actions.

The left-to-right degradation constraint is enforced through a combination of projection and sorting mechanisms. First, we apply a projection step that directly sets transition probabilities to zero for backward transitions: $p^a_{kk'} = 0$ for $k < k'$ for each action $a$. Second, we sort the emission parameters $\{\boldsymbol{\mu}_k\}$ across all actions to maintain a consistent ordering where $\boldsymbol{\mu}_1 \leq \boldsymbol{\mu}_2 \leq \ldots \leq \boldsymbol{\mu}_K$ in terms of their magnitude or a chosen feature dimension. This dual approach ensures both mathematical constraints on transition probabilities and physical ordering of degradation states, providing robust enforcement of the left-to-right constraint.

In contrast, action-dependent emissions are necessary when actions significantly influence sensor readings even for the same underlying degradation state. For example, in aviation applications, engine emissions, vibration patterns, and sensor readings differ substantially for different flight speeds or altitudes even for identical engine health conditions. In such cases, the emission model should be $P(\mathbf{O}_t | S_t=j, A_t=a)$ with action-specific parameters.

Our choice of shared emission parameters is justified by the physical nature of our degradation system, where the fundamental sensor signatures of each degradation state remain consistent across different operating speeds, while only the transition rates between states vary with actions.

\subsubsection{Parameter Estimation: Generalized Expectation-Maximization \label{subsub:GEM}}
Following the IOHMM framework, the model is trained using the generalized expectation-maximization algorithm:

\textbf{Initialization.}
For each action $a\in\mathbf{A}$ we initialize
the emission and transition parameters as follows:
(i) apply $k$--means clustering to the feature vectors
$\{\mathbf{O}_{i,t}\}$ corresponding to action $a$
to obtain initial emission means $\{\mu_k^{(a)}\}$;
(ii) set the covariances
$\Sigma_k^{(a)}$ to the within--cluster covariances
with a small ridge term $\varepsilon I$ ($\varepsilon=10^{-6}$)
to ensure positive definiteness; A left--to--right constraint is enforced by projecting
backward transitions to zero and sorting
the emission means $\{\mu_k^{(a)}\}$ across all actions
to maintain a consistent state order.

\textbf{E--step.}
This step follows the standard EM structure,
but the forward--backward recursions are
\emph{input--dependent} through the action sequence $\{A_t\}$.
The forward variable is updated as
\[
\alpha_{k,t}
= P\!\left(\mathbf{O}_t \mid S_t = k, A_t \right)
  \sum_{\ell=1}^K
  P\!\left(S_t = k \mid S_{t-1} = \ell, A_t \right)
  \alpha_{\ell,t-1},
\]
where $k$ and $\ell$ denote the next and current states, respectively.
The backward variable $\beta_{k,t}$ is computed analogously.
From these messages we obtain the posterior responsibilities
\[
\hat g_{k,t}
  = P\!\left(S_t = k \mid \mathcal{D}\right), \qquad
\hat h_{k\ell,t}
  = P\!\left(S_t = k,\; S_{t-1} = \ell \mid \mathcal{D}\right),
\]
where $\mathcal{D}$ denotes the entire observation sequence.

\textbf{M--step.}
Given the posteriors $\hat g_{k,t}$ and $\hat h_{k\ell,t}$
from the E--step, the model parameters are updated as follows.

\emph{Transition matrices.}
For each action $a\in\mathbf{A}$,
the input--dependent transition matrix
$\mathbf{A}^{(a)}\in\mathbb{R}^{K\times K}$ is updated by
\[
\bigl[\mathbf{A}^{(a)}\bigr]_{k\ell}
=\frac{\displaystyle
      \sum_{p=1}^{P}\sum_{t=1}^{T_p}
      \hat h_{p,k\ell}(t)\,\mathbf{1}(A_t=a)}
      {\displaystyle
      \sum_{p=1}^{P}\sum_{t=1}^{T_p}
      \sum_{r=1}^{K}
      \hat h_{p,r\ell}(t)\,\mathbf{1}(A_t=a)},
\]
where $\mathbf{1}(A_t=a)$ equals 1 if the action at time $t$ is $a$
and 0 otherwise.

\emph{Emission parameters.}
The updates depend on whether the emissions are shared
or action--dependent:
\begin{itemize}
\item \textbf{Shared emission (Case 1).}
\[
\mu_k=
\frac{\sum_{i,t}\hat g_{i,k}(t)\,\mathbf{O}_{i,t}}
     {\sum_{i,t}\hat g_{i,k}(t)},
\qquad
\Sigma_k=
\frac{\sum_{i,t}\hat g_{i,k}(t)
      (\mathbf{O}_{i,t}-\mu_k)(\mathbf{O}_{i,t}-\mu_k)^\top}
     {\sum_{i,t}\hat g_{i,k}(t)} .
\]

\item \textbf{Action--dependent emission (Case 2).}
For each action $a$,
\[
\mu_k^{(a)}=
\frac{\sum_{i,t}\hat g_{i,k}^{(a)}(t)
      \mathbf{O}_{i,t}\,\mathbf{1}(A_t=a)}
     {\sum_{i,t}\hat g_{i,k}^{(a)}(t)\,\mathbf{1}(A_t=a)},
\quad
\Sigma_k^{(a)}=
\frac{\sum_{i,t}\hat g_{i,k}^{(a)}(t)
      (\mathbf{O}_{i,t}-\mu_k^{(a)})
      (\mathbf{O}_{i,t}-\mu_k^{(a)})^\top
      \mathbf{1}(A_t=a)}
     {\sum_{i,t}\hat g_{i,k}^{(a)}(t)\,\mathbf{1}(A_t=a)} .
\]
\end{itemize}

\begin{algorithm}
\SetAlgoLined
\KwIn{Number of states $K$, multi-sensor observations $\{\mathbf{O}_{i,t}^a\}$, actions $\{a_t\}$}
\KwOut{Trained transition matrices $\mathbf{A}^{(a)}$, emission parameters $\{\boldsymbol{\mu}_k, \boldsymbol{\Sigma}_k\}$}
\SetKwFunction{FMain}{IOHMM Training}
\SetKwProg{Fn}{Function}{:}{}
\Fn{\FMain{}}{

    Initialize transition matrices $\mathbf{A}^{(a)}$ with left-to-right constraint;
    Initialize emission parameters $\{\boldsymbol{\mu}_k, \boldsymbol{\Sigma}_k\}$ using k-means; 

     \While{Not converged}{
            E-Step: Compute $\hat{g}_{i,t}$, $\hat{h}_{ij,t}$ using input-dependent forward-backward algorithm;
        
            M-Step: Update transition matrices for each action $a$;
            $[\mathbf{A}^{(a)}]_{ij} \leftarrow \frac{\sum_{p,t} \hat{h}_{ij,t} \cdot \mathbf{1}(A_t = a)}{\sum_{p,t,k} \hat{h}_{kj,t} \cdot \mathbf{1}(A_t = a)}$;
            
            Apply projection constraint: $[\mathbf{A}^{(a)}]_{ij} = 0$ for $i < j$ and normalize rows;
            
            Update emission parameters using standard EM updates;
            $\boldsymbol{\mu}_k, \boldsymbol{\Sigma}_k \longleftarrow \text{EM-update}(\hat{g}_{i,t}, \mathbf{O}_{i,t})$;
            
            Enforce left-to-right constraint: Sort emission parameters $\{\boldsymbol{\mu}_k\}$ to ensure $\boldsymbol{\mu}_1 \leq \boldsymbol{\mu}_2 \leq \ldots \leq \boldsymbol{\mu}_K$;
            Reassign state indices based on sorted emission parameters and update corresponding transition matrices for all actions;
        }

}
\textbf{end}
\caption{IOHMM training using Generalized EM algorithm}
\end{algorithm}

\textbf{Combined Projection and Sorting Mechanism for Left-to-Right Constraint:}
The left-to-right constraint is enforced through two complementary steps that ensure both mathematical constraints and physical consistency:

\textbf{Step 1: Projection Constraint}
Apply direct mathematical constraint by setting \(A^{(a)}_{k k'} = 0 \; (\forall\, k'<k)\)
 and normalizing the remaining probabilities to ensure valid transition matrices.

\textbf{Step 2: Sorting Constraint}
For Shared Emission Matrix:
1. After updating emission parameters $\{\boldsymbol{\mu}_k, \boldsymbol{\Sigma}_k\}$, sort them based on a chosen feature dimension.
2. Let $\sigma$ be the permutation that sorts $\{\boldsymbol{\mu}_k\}$ in ascending order.
3. Reassign state indices: $\boldsymbol{\mu}_k^{new} = \boldsymbol{\mu}_{\sigma(k)}$ and $\boldsymbol{\Sigma}_k^{new} = \boldsymbol{\Sigma}_{\sigma(k)}$.
4. Update transition matrices: $[\mathbf{A}^{(a)}]_{ij}^{new} = [\mathbf{A}^{(a)}]_{\sigma^{-1}(i),\sigma^{-1}(j)}$ for all actions $a$.


This dual approach ensures both mathematical constraints on transition probabilities (through projection) and physical consistency in state ordering (through sorting), providing robust enforcement of the left-to-right constraint that is both theoretically sound and practically effective.

\subsubsection{Continuous-to-Discrete Observation Mapping for POMDP Compatibility}
\label{subsub:discrete}
While sensor data $\mathbf{O}$ is naturally continuous, POMDP solvers typically require discrete observation spaces $\Omega$ for computational tractability. This discretization step bridges the gap between continuous sensor readings and the discrete POMDP framework while preserving the essential information needed for decision-making.

The key motivation for discretization stems from three practical considerations: (1) \textbf{Computational Efficiency}: Continuous observation spaces lead to intractable belief state updates and policy optimization, as the belief space becomes infinite-dimensional; (2) \textbf{Sufficient Resolution}: Discrete observations with appropriate granularity provide adequate information for maintenance decisions without sacrificing policy quality; and (3) \textbf{Standard POMDP Algorithms}: Most established POMDP solution algorithms (e.g., PBVI, HSVI) are designed for discrete observation spaces and scale poorly with continuous observations.

We transform continuous sensor features $\mathbf{O}$ into discrete observations using a Gaussian Mixture Model (GMM) approach. The GMM consists of $k$ components, where each component $i$ represents a discrete observation symbol $o_i$ with parameters $(\phi_i, \mu_i, \sigma_i)$ denoting the weight, mean, and variance respectively. The mapping is performed by computing the posterior probability (responsibility) of each discrete symbol given the continuous observation: $\gamma_i = P(o_i \mid \mathbf{O})$, and assigning the observation to the most probable symbol: $\mathbf{o} = \arg\max_i P(o_i \mid \mathbf{O})$. This approach leverages the probabilistic nature of the GMM to provide soft assignment capabilities while maintaining computational efficiency. Alternative clustering algorithms (e.g., k-means, hierarchical clustering) can be employed, but we observe minimal performance differences across different clustering methods, suggesting that the choice of discretization algorithm is not critical for policy quality. 

The posterior probability (responsibility) can be calculated as following, where the model parameters can be estimated through expectation maximization:

\begin{equation}
\label{eq:gmm_posterior}
\begin{aligned}
P(o_i \mid \mathbf{O})
&= \frac{P(o_i, \mathbf{O})}{P(\mathbf{O})}
 = \frac{P(o_i) \, P(\mathbf{O} \mid o_i)}
        {\sum_{j=1}^{k} P(o_j) \, P(\mathbf{O} \mid o_j)} \\
&= \frac{\phi_i \, \mathcal{N}(\mathbf{O} \mid \mu_i, \sigma_i)}
        {\sum_{j=1}^{k} \phi_j \, \mathcal{N}(\mathbf{O} \mid \mu_j, \sigma_j)} .
\end{aligned}
\end{equation}

With the IOHMM parameters successfully estimated, we now have a fully specified POMDP model with known transition probabilities $T$ and observation probabilities $O$. The final step is to solve this POMDP to obtain an optimal policy and implement it in real-time, which is the focus of Section \ref{subsec:pomdp_solution}.

\begin{algorithm}
\SetAlgoLined
\KwIn{Observing signal $z$, $\phi, \mu, \sigma$}
\KwOut{Belief space $\mathbf{b}$}

    \SetKwFunction{FMain}{GMM}
    \SetKwProg{Fn}{Function}{:}{}
    \Fn{\FMain{}}{
            
        $O \longleftarrow \{X_{rms}(z), \ldots, X_{e}(z)\}$;//Abstract features from signals using expressions in Table 2 
        
        $\mathbf{b(o|z)} \longleftarrow$ $\frac{\phi_i\mathcal{N}(\mathbf{O}|\mu_i,\sigma_i)}{\sum_{j=1}^k\phi_j\mathcal{N}(\mathbf{O}|\mu_j,\sigma_j)}$; // Obtain the belief on the continuous observation over the discrete observations

    }
    \textbf{end}

 \caption{GMM Classifier}
\end{algorithm}





as the probability distribution (refer to Section 2.1 for solution methodology) that predicts the observation outcome when the system is in state $S'$ following action $a_t$. This equation is central to our decoding challenge. The evolution of the belief state to $S'$, contingent on action $a$ and observation $O$, is described in the following state transition equation:

\begin{equation}
    b_o^a(S') = \frac{Z(S', a, O)\sum \limits_{S \in \mathcal{S}}X(S, a, S')b(S)}{\sum\limits_{S' \in \mathcal{S}}\Big \{Z(S', a, O)\sum \limits_{S \in \mathcal{S}}X(S, a, S')b(S)\Big \}}
\end{equation}

\paragraph{Reward Function}
Continuing, we introduce $J_\pi(b_0)$, representing the anticipated cumulative discounted reward, given initial belief state $b_0$ and adhering to policy $\pi \in \boldsymbol{\pi}$. This is expressed as

\begin{equation}
\begin{aligned}
& \underset {\boldsymbol {\pi}}{\text{Maximize}}
& & J_{\pi}(b) = \mathbf{E}[ \sum\limits_{t=1}^{\infty}\revised{\gamma^t} R(b,\pi)|b_0],\\
\end{aligned}
\label{eq_obj_pj1}
\end{equation}
where 
\[
R(b,a) = \sum_{S \in \mathcal{S}} b(S)\,r(S,a),
\]
and $\revised{\gamma}$ is the discount factor. Here, $\pi$ encapsulates all decisions relevant to the PM policies.

To describe the expected total reward achievable from period $t$ to infinity, given the belief state at time $t$, we use $V_t(b)$ with $V_0(\cdot) = 0$. The following equation captures this concept:

\begin{equation}
\label{eq_bellman_pj1}
V_t(b)
\,=\, \max_{a\in\boldsymbol{\pi}}
\Bigg\{
  R(b,a)
  + \gamma \sum_{o\in\Omega}
  \Pr(O \mid a,b)\,
  V_{t+1}(b_{o}^a)
\Bigg\}.
\end{equation}

in which 
\[Pr(O|a,b)=\sum\limits_{S' \in \mathcal{S}}\Big \{Z(S', a, O)\sum \limits_{S \in \mathcal{S}}X(S, a, S')b(S)\Big \}.\]

The latter part of Equation (\ref{eq_bellman_pj2}) clearly delineates two distinct aspects of the reward calculation: firstly, the immediate reward associated with the current state at time t, and secondly, the anticipated future rewards, adjusted for time-value, for all potential future states of the system commencing from the subsequent time period. This framework facilitates the computation of the most effective strategies from the initial time point, employing the value iteration method as detailed by Puterman (2005) \cite{puterman2014markov}. To derive both the optimal policy and its corresponding value, we implement the Point-Based Value Iteration (PBVI) algorithm, as proposed by Pineau, Gordon, and Thrun (2003) \cite{pineau2005point}. This algorithm is pivotal in generating a collection of $alpha$-vectors, corresponding to a selected set of belief points. This approach is instrumental in formulating the optimal solutions within our context. 

\begin{algorithm}
\SetAlgoLined
\KwIn{
A POMDP tuple $(h_S, X, tr, R, \Omega, O, b_0)$}
\KwOut{$\alpha$, $B$}
    \SetKwFunction{FMain}{PBVI}
    \SetKwProg{Fn}{Function}{:}{}
    \Fn{\FMain{}}{
            
        $B \longleftarrow \{b_0\}$\;   
        
        \While{Not converged}{
          Improve(V,B)
          
          B $\longleftarrow$ Expand(B)
          
        }
    }
    \textbf{end}

    \SetKwFunction{FImprove}{Improve}
    \Fn{\FImprove{$V,B$}}{
           
        \While{Not converged}{

            \ForEach{$ b \in B $}
            {$\alpha \longleftarrow$ Backup(b,V)
            
            $ V \longleftarrow V \cup  \{\alpha \}$
            
            }
            
        }
    }
            
    \textbf{end}
    
    \SetKwFunction{FExpand}{Expand}
    \Fn{\FExpand{$B$}}{
        $ B'  \longleftarrow B $;    

        \ForEach{$ b \in B $}
        {$\Gamma(b)\longleftarrow\{b^{a,o}|Pr(o|b,a)>0\}$
        
        $B'\longleftarrow B'\cup argmax_{b'\in \Gamma(b)}~\|B,b'\|_2$
        
        }
        \textbf{return} B'
    }
    \textbf{end}

    \SetKwFunction{FBackup}{Backup}
    \Fn{\FBackup{$b,V$}}{

        \ForEach{$ b_0 \in b $}
        {$\beta_{a,O} \longleftarrow argmax_{\alpha \in V}~\alpha b_o^a$
        
       $ \beta_{a}(s)                   
 \longleftarrow R(s,a) +\revised{\gamma}\sum\limits_{S}\sum\limits_{S'} Z(S',a,O)X(S,a,S')\beta_{a,O}$
        
        $\beta \longleftarrow argmax_{\beta_a} ~\beta_a b$
        
        }
        
        \textbf{return} $\beta$
        
    }
    
    \textbf{end}

 \caption{PBVI Algorithm to solve POMDP}
\end{algorithm}
\subsection{Optimal Policy Generation and Real-Time Implementation \label{subsec:pomdp_solution}}
This final section describes how to use the fully specified model to find and apply the optimal policy. With the parameters estimated from the IOHMM framework, the POMDP is now fully defined and ready for solution.

\subsubsection{Offline Policy Generation \label{subsub:offline_policy}}
With the parameters estimated from the IOHMM, the POMDP is now fully defined and ready for solution. Before introducing the solution algorithm, we first establish the concept of belief states, which is fundamental to POMDP solution methods.

\paragraph{Belief State Concept}
The POMDP framework introduces the concept of a belief state $b(s)$, which represents the probability distribution over the hidden states given the history of observations. Since the true system state is not directly observable, the belief state serves as a sufficient statistic that encapsulates all relevant information for decision-making. The objective is to solve the Bellman equation to maximize the expected discounted reward:

\begin{equation}
\label{eq_bellman_pj2}
V^*(b) = \max_{a \in A} \left[ R(b,a) + \gamma \sum_{O \in \Omega} Pr(O|a,b) V^*(b') \right]
\end{equation}

where $b'$ is the updated belief state after taking action $a$ and observing $o$.

In order to generate the optimal control policy, we use the Point-Based Value Iteration (PBVI) algorithm \cite{pineau2005point}. Specifically, we would like to estimate the actions $a_t$ at time $t$ of the POMDP problem defined in Section \ref{subsec:pomdp_formulation} assuming that parameters of the transition probability matrix $T$ are given. 

To solve the POMDP, we will define the following important concepts. 

\paragraph{Belief Update}
The belief state evolves over time as new observations are received. Let $Z(S',a, O') = Pr(O_{t+1} = O' | S_{t+1} = S',~a)$ be the probability of observing $O'$ given that the system is in state $S'$ and action $a$ is taken. The belief update equation is:

\begin{equation}
    b'(S') = \frac{Z(S',a, O') \sum_{S \in \mathcal{S}} X(S, a, S') b(S)}{Pr(O|a,b)}
\end{equation}

where $X(S, a, S')$ represents the transition probability from state $S$ to $S'$ under action $a$, and $Pr(O|a,b)$ is the probability of observing $O$ given action $a$ and belief state $b$.

\paragraph{Value function}
The value function for a belief state $b$ is defined as:

\begin{equation}
\label{eq_bellman_pj2_v2}
V^*(b) = \max_{a \in A} \left[ R(b,a) + \gamma \sum_{O \in \Omega} Pr(O|a,b) V^*(b') \right]
\end{equation}

in which 
\[Pr(O|a,b)=\sum\limits_{S' \in \mathcal{S}}\Big \{Z(S', a, O)\sum \limits_{S \in \mathcal{S}}X(S, a, S')b(S)\Big \}.\]

The latter part of Equation (\ref{eq_bellman_pj2}) clearly delineates two distinct aspects of the reward calculation: firstly, the immediate reward associated with the current state at time t, and secondly, the anticipated future rewards, adjusted for time-value, for all potential future states of the system commencing from the subsequent time period. This framework facilitates the computation of the most effective strategies from the initial time point, employing the value iteration method as detailed by Puterman (2005) \cite{puterman2014markov}. To derive both the optimal policy and its corresponding value, we implement the Point-Based Value Iteration (PBVI) algorithm, as proposed by Pineau, Gordon, and Thrun (2003) \cite{pineau2005point}. This algorithm is pivotal in generating a collection of $\alpha$-vectors, corresponding to a selected set of belief points. This approach is instrumental in formulating the optimal solutions within our context.

\subsubsection{Online Real-Time Control \label{subsub:online_control}}
Outline the step-by-step procedure for using the policy in real-time. This corresponds to the current "Obtain the actions for observing signals" section.

For a given set of observed signals, it will first be transformed to a set of discrete observations using the feature abstraction approach as discussed in Section 4.2. Then the GMM algorithm will deliver the set of estimated probabilities, which represents the belief that the observation matches with the predefined observations in the POMDP framework, i.e., $\mathbf{B(o)}=\{B(o_1),B(o_2),,\ldots, B(o_J)\}$. Then the belief on the hidden states of this observation can be obtained using the expressing $\mathbf{\Bar{b}(o)}=\mathbf{B(o)b^T(s)}$. The action that the machine should take under the observation $o$ then will be the one associated with the $\alpha$ vector that brings the maximum reward, which is expressed as:

\begin{equation}
	a = {\arg\max}_{a_i:i\in \{1,2,\ldots\,|\mathbf{\alpha}|\}}\ \ \mathbf{\alpha_i \mathbf{b}}(o)\mathbf{B^T}(o).
\end{equation}

\begin{algorithm}
\SetAlgoLined
\KwIn{Observing signal $z$, $\mathbf{\Lambda}$,$\mathbf{a}$, emission matrix $\mathbf{B}$}
\KwOut{Action a}
    \SetKwFunction{FMain}{Decision}
    \SetKwProg{Fn}{Function}{:}{}
    \Fn{\FMain{}}{
            
        
        $\mathbf{b'(o|z)} \longleftarrow$ \texttt{GMM}($z$); // Obtain the belief on the discrete observations using GMM algorithms
        
        $\mathbf{b(s|z)} \longleftarrow  \mathbf{Bb'(o|z)}$;// Get the hidden state belief
        
        $a \longleftarrow {\arg\min}_{a_l:l\in \{1,2,\ldots\,|\mathbf{\alpha}|\}}\ \ \mathbf{b^T(s|z)\alpha_l}.$
        
    }
    \textbf{end}

 \caption{Algorithm to obtain the action for observing signals}
\end{algorithm}

The overall algorithm is summarized as shown in Figure \ref{fig:algo_structure}. For the offline training, as shown on the left side, historical data sets are collected to abstract the features and thereafter used to obtain the parameters for the HMM model and the POMDP model, which will ultimately return the vectors for decision-making given different observations. When the model is conducted for real-time control, the signal is abstracted into the features, which will be classified as the belief vectors on the discrete observations. The belief vector, together with the alpha vectors generated from the training stage, will be used to determine the ultimate control action at the exact time.

\section{Case Studies}
\label{sec:case_studies}

This section presents comprehensive case studies to demonstrate the effectiveness and generalizability of the proposed constrained IOHMM framework for degradation modeling and maintenance optimization. We evaluate our methodology on two distinct datasets that represent different industrial applications and degradation mechanisms:

\begin{itemize}
    \item \textbf{XJTU-SY Bearing Dataset}: A laboratory experimental platform with multiple operating conditions, featuring vibration-based degradation monitoring of rolling element bearings under controlled loading and speed variations.
    \item \textbf{NASA C-MAPSS Turbofan Dataset}: A widely-used benchmark dataset for prognostics, featuring multi-sensor monitoring of aircraft engine degradation under operational conditions.
\end{itemize}

Each case study follows a consistent structure: (1) dataset introduction and characteristics (Sections \ref{subsec:xjtu_dataset_intro} and \ref{subsec:nasa_dataset_intro}), (2) feature extraction and parameter estimation (Sections \ref{subsec:xjtu_feature_extraction} and \ref{subsec:nasa_feature_extraction}), and (3) results analysis including degradation modeling, RUL prediction, and maintenance policy optimization (Sections \ref{subsec:xjtu_results_analysis} and \ref{subsec:nasa_results_analysis}). This systematic approach allows for comprehensive evaluation of the proposed framework across different domains and validates its applicability to real-world industrial scenarios. Following the case studies, Section \ref{sec:results_sensitivity_analysis} provides detailed sensitivity analysis, model selection criteria, and comparison with baseline methods to demonstrate the robustness and effectiveness of the proposed approach.

\subsection{Case Study 1: XJTU-SY Bearing Dataset}
\label{subsec:case_study_1}

In this case study, we evaluate the proposed algorithm using a real bearing dataset from a laboratory experimental platform \cite{wang2018hybrid} with multiple operating conditions. The bearing degradation experiments provide controlled conditions to validate our IOHMM framework's ability to capture condition-dependent degradation dynamics and generate optimal maintenance policies. 

\subsubsection{Dataset Introduction}
\label{subsec:xjtu_dataset_intro}

To validate the effectiveness of the proposed framework, we use the XJTU-SY bearing dataset developed from a laboratory experimental platform (\cite{wang2018hybrid}). Accelerated degradation tests of bearings
conducted on the platform, provide data that characterizes the bearing degradation process during the entire operating lifetime.   The testing bed of bearing degradation experiments consists of an alternating current (AC) induction motor, a support shaft, two support roller bearings, a hydraulic loading system, and a motor speed controller. The acceleration degradation test is conducted for bearings under different operation conditions, typically different rotating speeds and radial forces, which can be controlled by the hydraulic loading system and the speed controller of the AC induction motor. The experimental operating conditions are listed in Table \ref{tab:parameter}, as well as the performance of each condition, such as the lifetime and the Maximum Amplitude (MA). There are in total three different operating conditions, and brand-new bearings (model LDK UER204) were tested for each operating condition.


The vibration signals from the horizontal direction, which also matches the loading direction, are captured to track the degradation process. The signals are collected with a sampling frequency of 25.6 kHz (32768 data points per minute). The degradation process is therefore monitored and forced to stop when the MA values of the signal reach out of the predetermined safety threshold, which is considered machine failure. 

\begin{figure}%
    \centering
    \subfloat[]{\includegraphics[width=0.33\linewidth]{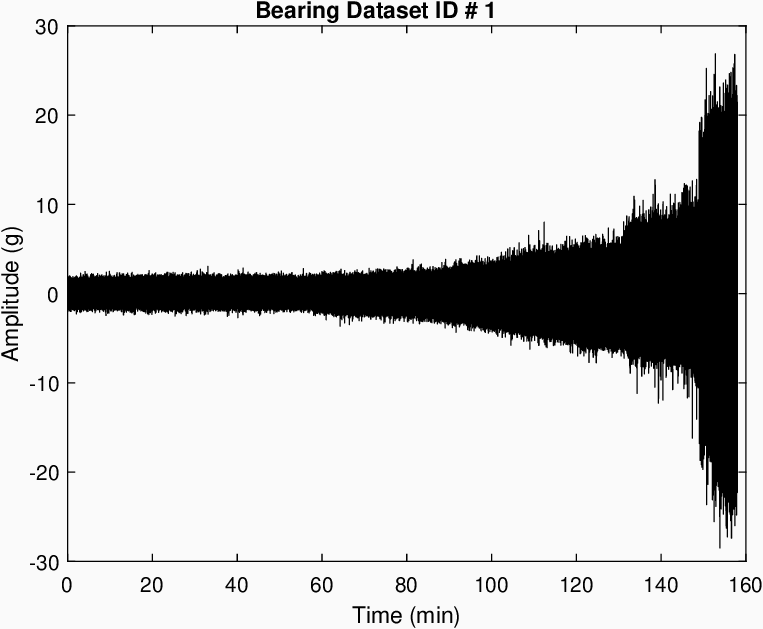}}%
    \subfloat[]{\includegraphics[width=0.33\linewidth]{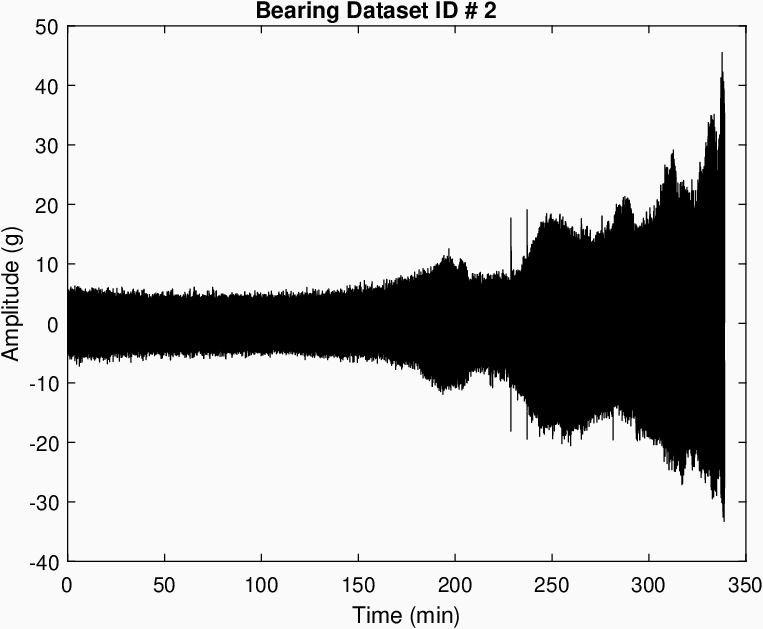}}%
    \subfloat[]{\includegraphics[width=0.33\linewidth]{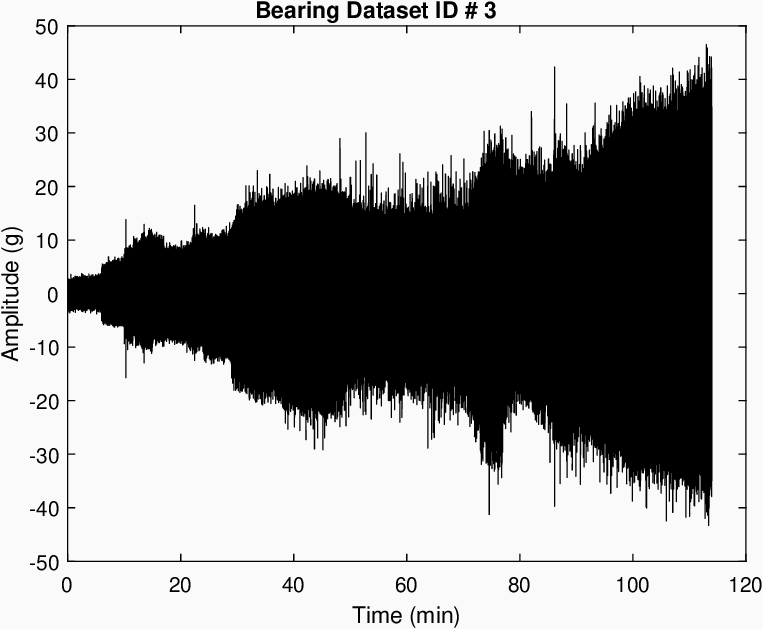}}%
    \caption{The plot of vibration signals under different operating conditions}%
    \label{fig:example}%
\end{figure}

\begin{table}[]
    \centering
      \caption{Information about the bearing degradation dataset}
    \begin{tabular}{c|ccc}
    \hline
         ID& Operating Condition& Bearing Lifetime & Max MA \\
         \hline
         1& 35Hz/12kN & 158 min & 28 g\\
         2& 37.5Hz/11kN & 339 min & 45 g \\
         3& 40Hz/10kN & 114 min & 45 g\\
         \hline
    \end{tabular}
  
    \label{tab:parameter}
\end{table}
Also, the degradation plot to the dataset under different operating conditions is shown in Figure \ref{fig:example}. Overall, it can be found that an increasing trend is presented in the amplitudes of the vibration signals throughout the entire operating time. For some operating conditions, i.e., as shown in sub-figure (a) and (b), a fluctuation in the signal values occurs before the fast increase of the amplitudes.

\subsubsection{Feature Extraction and Parameter Estimation}
\label{subsec:xjtu_feature_extraction}

This subsection presents the feature extraction methodology and parameter estimation process for the XJTU-SY bearing dataset. The analysis involves two main components: (1) extracting statistical features from raw vibration signals to capture degradation characteristics, and (2) applying the constrained IOHMM framework to estimate model parameters and learn the degradation dynamics. The feature extraction process transforms high-frequency vibration data into meaningful statistical descriptors that characterize the bearing's health state, while the parameter estimation phase determines the optimal model parameters that best capture the observed degradation patterns.

For the process to derive the degradation signals from the raw degradation data (i.e., the vibration data), the following consecutive steps are commonly applied. 
Eleven statistical features are extracted from the vibration signals, i.e., root mean square, mean value, standard deviation, skewness, kurtosis, shape factor, peak to peak, energy, etc. The expressions of the features are listed out as shown in Table \ref{feature}, in which $x_i$ is the $i^{th}$ sample of the timed signal, and $N$ is the total number of the samples. These features indicate the amplitude and reflect the distribution of a signal over the time domain. 
\begin{table}[h]
    \centering
     \caption{Feature descriptions of the dataset}
\begingroup
\setlength{\tabcolsep}{10pt} 
\renewcommand{\arraystretch}{1.5} 
\resizebox{\columnwidth}{!}{%
    \begin{tabular}{l c l}
 \toprule
         Feature&~~~~~~~~~~~~~& Expression\\
         \hline
RMS & &$X_{rms} = (\frac{1}{N}\sum\limits_{i=1}^{N}x_i^2)^{\frac{1}{2}}$ \\
Mean& &$X_{m} =\frac{1}{N}\sum\limits_{i=1}^{N}x_i$\\
Standard deviation (Std) & &$X_{std} =(\frac{1}{N}\sum\limits_{i=1}^{N}(x_i-X_{m})^2)^{\frac{1}{2}}$\\
Skewness& &$X_{ss} =\frac{1}{X_{rms}^3}\sum\limits_{i=1}^{N}(x_i-X_{m})^3$ \\
Kurtosis &&$X_{ks} =\frac{1}{X_{rms}^4}\sum\limits_{i=1}^{N}(x_i-X_{m})^4$\\
Peak to peak (P-P) &&$X_{pp} =x_{max}-x_{min}$ \\
Crest factor &&$X_{cf} =\frac{x_{max}}{X_{rms}}$\\
Shape factor &&$X_{sf} =\frac{x_{rms}}{\frac{1}{N}\sum\limits_{i=1}^{N}|x_i|}$ \\
Impulse factor &&$X_{if} =\frac{x_{max}}{\frac{1}{N}\sum\limits_{i=1}^{N}|x_i|}$ \\
Margin factor &&$X_{mf} =\frac{x_{max}}{(\frac{1}{N}\sum\limits_{i=1}^{N}|x_i|)^2}$ \\
Energy &&$X_{e} =\sum\limits_{i=1}^{N}x_i^2$ \\
  \bottomrule
    \end{tabular}
    }
\endgroup
   
    \label{feature}
\end{table}

\begin{figure}
    \centering
    \includegraphics[width=1\linewidth]{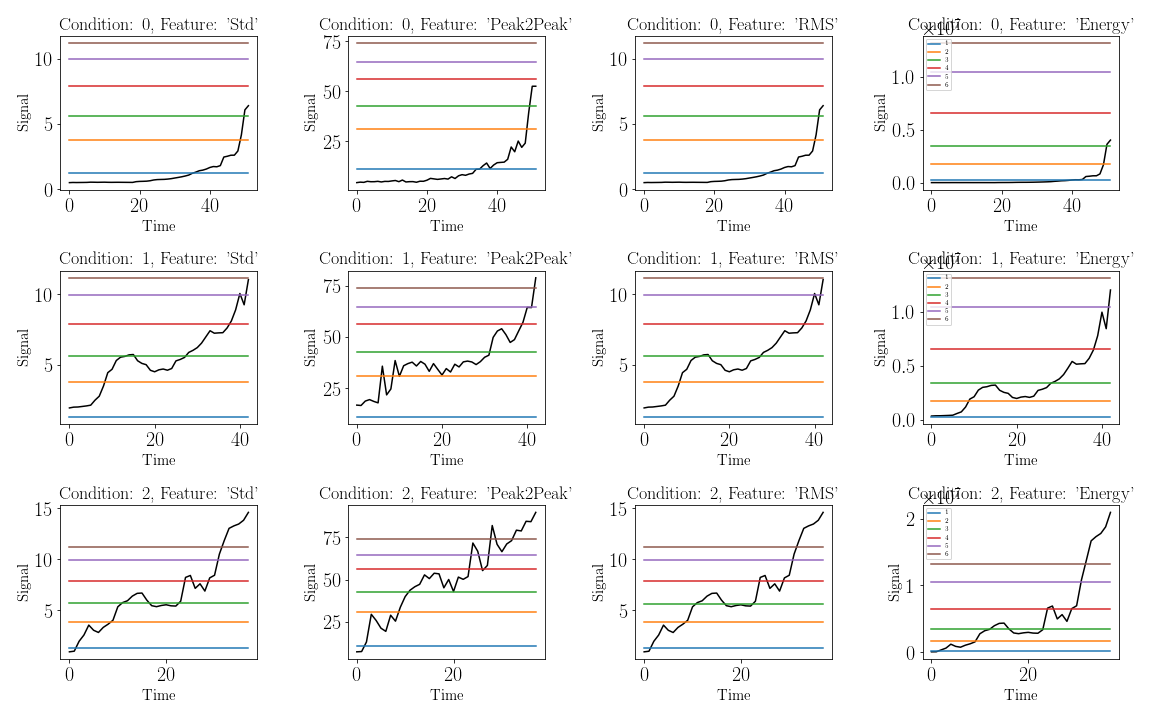}
    \caption{State estimation for all conditions}
    \label{fig:6_state}
\end{figure}

\begin{figure}
    \centering
    \includegraphics[width=1\linewidth]{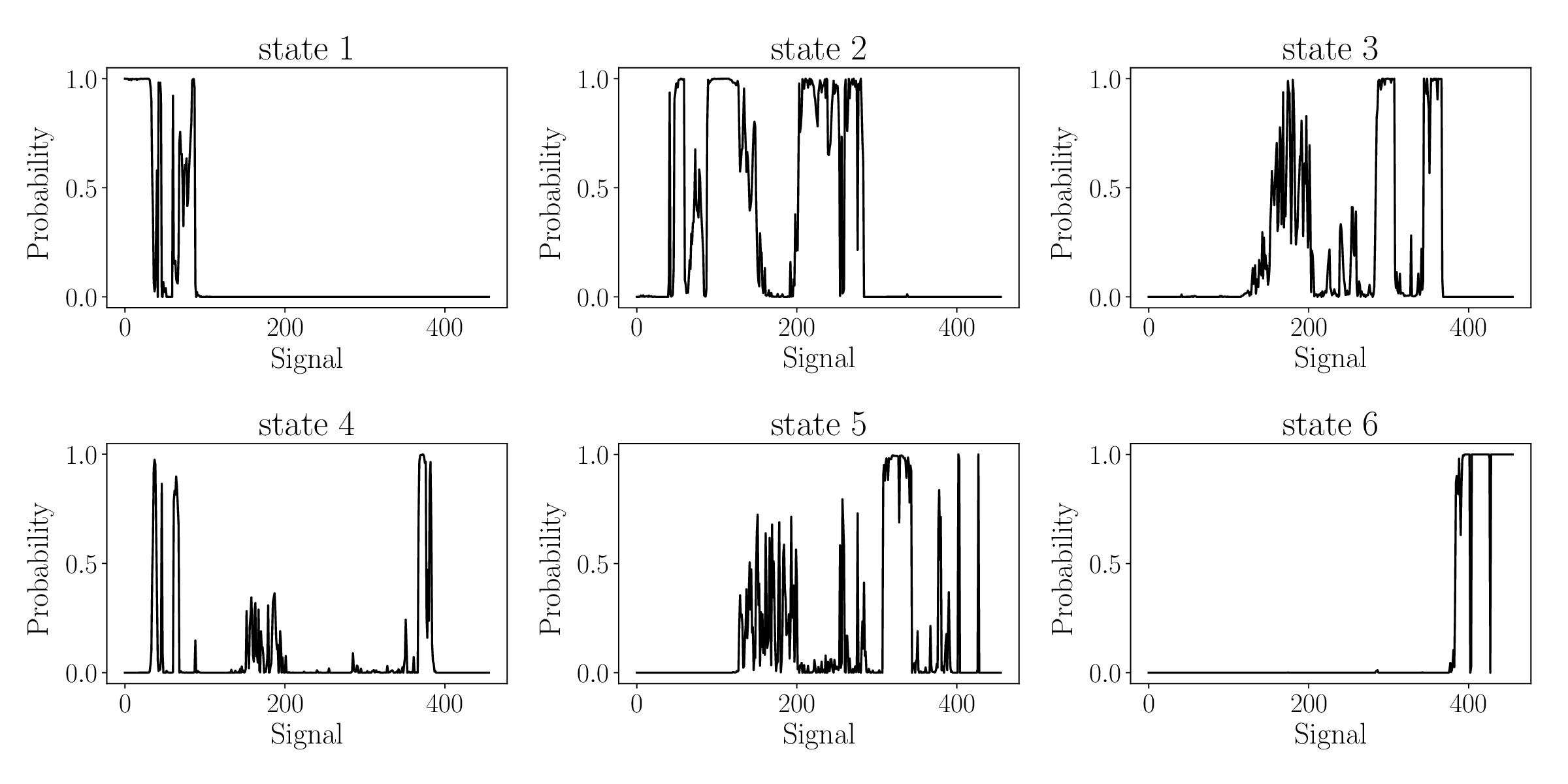}
    \caption{Posterior probability for the hidden states}
    \label{fig:posterior}
\end{figure}

\textbf{Parameter Estimation Results.} We then apply the proposed constrained-IOHMM models to classify the original signal into six hidden unobservable states, with the last state representing the failure observable state. In Figure \ref{fig:6_state}, four representing features are selected under the three operating conditions. The values of these features are plotted under each operation condition, with the colored bars representing the level of the mean values under different hidden states. It can be found that the value of all these features indicates an upward trend overwhelmingly, which is consistent with the trend of the original signal as well. Such behavior is also adaptive to a left-to-right constraint in the IOHMM model. Besides, the range of these features is also different under different conditions. For example, under the feature "Std", the signal value is below 10 for condition 0 (lower capacity) while it surpasses 15 for condition 3 (higher capacity). Therefore, for the lower capacity case, the system has less probability to retain the inferior hidden states than the other conditions before the failure of the machine is observed.  Furthermore, the posterior probabilities for the hidden states are shown in Figure \ref{fig:posterior}. From the results, we can see that the system is changing states gradually. The first state and last state are able to capture the initial and failure signal pretty clear. However, there is much more noise in the middle states. For example, state 4 is related to signals close to both start and end. The reason is that we assume that multiple conditions will share the same emission matrix. Thus, one state might be corresponding to signals from a different time.

\subsubsection{Results and Analysis}
\label{subsec:xjtu_results_analysis}

This subsection presents comprehensive analysis of the experimental results obtained from applying the constrained IOHMM framework to the XJTU-SY bearing dataset. The analysis encompasses three key aspects: (1) RUL prediction performance evaluation, which demonstrates the framework's ability to forecast remaining useful life with quantified uncertainty; (2) maintenance policy optimization, which shows how the learned degradation model enables optimal decision-making for capacity adjustment and preventive maintenance; and (3) interpretation of the learned model parameters and their physical significance. The results validate the effectiveness of the proposed framework in capturing degradation dynamics and generating actionable maintenance policies for real-world bearing systems.

\revised{
\textbf{RUL Prediction.} To validate the remaining useful life (RUL) prediction capability of the proposed constrained IOHMM framework, we conducted experiments comparing its forecasts with the true degradation trajectory. Figure \ref{fig:rul_prediction} plots the true RUL (red line), the median RUL predicted by the model (black solid line), and the 95 \% confidence interval (black dashed lines).

The true RUL decreases approximately linearly as time progresses, while the IOHMM prediction tracks the overall downward trend with several stepwise updates that reflect discrete latent-state transitions. Throughout the degradation process the predicted median remains within the confidence band and anticipates critical drops in RUL several decision cycles before a conventional amplitude-threshold detector would signal failure.

The widening and narrowing of the confidence interval capture the evolving uncertainty in the hidden state estimates, with narrower bounds as the system approaches failure.
These results demonstrate that the IOHMM not only follows the long-term degradation pattern but also provides statistically reliable early warnings of impending failure, enabling proactive maintenance actions before visible deviations appear in the raw vibration signals.

\begin{figure}[h]
    \centering
    \includegraphics[width=0.85\linewidth]{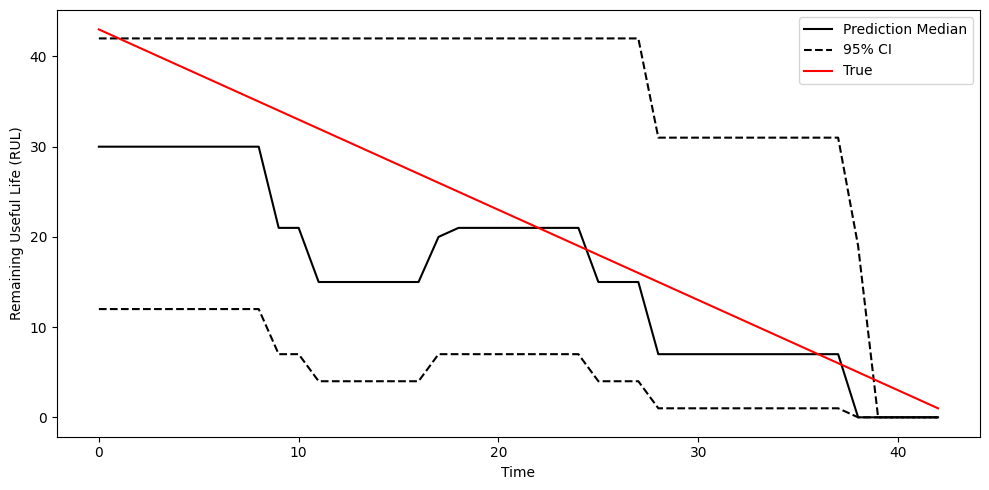}
    \caption{Constrained IOHMM-based RUL prediction compared with the true RUL.}
    \label{fig:rul_prediction}
\end{figure}
}

\textbf{Maintenance Policy Optimization.} We then consider the control of preventative maintenance and optimize the three operating conditions. Based on the estimated six latent states, we can generate the PM policies for the engine. The objective is to find best of the combination of the capacities and PM such that the total long-term production cost is the minimum. The state transition matrix for all the operating conditions can be expressed as follows:

\begin{equation}
\mathbf{A_1}=\begin{bmatrix}
    1 & 2 & 3 & 4 & 5 & F \\
    0.9330 & 0.0670 & 0 & 0 & 0 & 0 \\
    0 & 0.8735 & 0.1265 & 0 & 0 & 0 \\
    0 & 0 & 0.4639 & 0.5361 & 0 & 0 \\
    0 & 0 & 0 & 0.0844 & 0.9156 & 0 \\
    0 & 0 & 0 & 0 & 0.1091 & 0.8909 \\
    0 & 0 & 0 & 0 & 0 & 1
\end{bmatrix}
\end{equation}

\begin{equation}
\mathbf{A_2}=\begin{bmatrix}
    1 & 2 & 3 & 4 & 5 & F \\
    0.8119 & 0.1881 & 0 & 0 & 0 & 0 \\
    0 & 0.0428 & 0.9572 & 0 & 0 & 0 \\
    0 & 0 & 0.9048 & 0.0952 & 0 & 0 \\
    0 & 0 & 0 & 0.7707 & 0.2293 & 0 \\
    0 & 0 & 0 & 0 & 0.6425 & 0.3575 \\
    0 & 0 & 0 & 0 & 0 & 1
\end{bmatrix}
\end{equation}
\begin{equation}
\mathbf{A_3}=\begin{bmatrix}
    1 & 2 & 3 & 4 & 5 & F \\
    0.5606 & 0.4394 & 0 & 0 & 0 & 0 \\
    0 & 0.4445 & 0.5555 & 0 & 0 & 0 \\
    0 & 0 & 0.8823 & 0.1177 & 0 & 0 \\
    0 & 0 & 0 & 0.5407 & 0.4593 & 0 \\
    0 & 0 & 0 & 0 & 0.7195 & 0.2805 \\
    0 & 0 & 0 & 0 & 0 & 1
\end{bmatrix}
\end{equation}
The emission matrix is listed out as follows:
\begin{equation}
\mathbf{B}=\begin{bmatrix}
    1 & 2 & 3 & 4 & 5 \\
    0.1983 & 0.0000 & 0.8013 & 0.0000 & 0.0003 \\
    0.0025 & 0.0098 & 0.5522 & 0.0000 & 0.4355 \\
    0.0000 & 0.2810 & 0.1381 & 0.0000 & 0.5809 \\
    0.0000 & 0.7224 & 0.0279 & 0.0014 & 0.2359 \\
    0.0000 & 0.7531 & 0.0001 & 0.1630 & 0.0838 \\
    0.0000 & 0.7651 & 0.0000 & 0.1907 & 0.0442
\end{bmatrix}
\end{equation}

Besides, the cost matrix can be found out as follows:

\begin{equation}
\mathbf{c}=\begin{bmatrix}
\text{Action/State} & 1 & 2 & 3 & 4 & 5 & F \\
1 & 1.2 & 1.2 & 1.2 & 1.2 & 1.2 & -25 \\
2 & 1.3 & 1.3 & 1.3 & 1.3 & 1.3 & -25 \\
3 & 1.5 & 1.5 & 1.5 & 1.5 & 1.5 & -25 \\
\text{PM} & -6 & -6 & -6 & -6 & -6 & -25
\end{bmatrix}
\end{equation}

By solving the POMDP problem, we can obtain the collection of belief vectors (The $\alpha$ vectors in the algorithm) and the corresponding actions ($a$). These belief vectors are the sampled points, which partitioned the entire belief space. Table \ref{tab:belief_actions} lists out 5 of the belief vectors obtained by the PBVI algorithm.

\begin{table}[]
\centering
  \caption{A list of belief vectors and actions}
\resizebox{\columnwidth}{!}{%
\begin{tabular}{@{}c|ccccccc@{}}
\toprule
\multirow{2}{*}{Belief} & \multirow{2}{*}{Action} & \multicolumn{6}{c}{Value function}                       \\ \cmidrule(l){3-8} 
                        &                         & State 1 & State 2 & State 3 & State 4 & State 5 & F      \\ \midrule
1                       & C=1.2                   & 6.32    & 5.58    & 4.24    & -8.79   & -27.3   & -84.16 \\
2                       & C=1.2                   & 6.02    & 4.52    & 4.26    & 1.08    & -11.7   & -55.55 \\
3                       & C=1.5                   & 5.27    & 2.8     & 2.68    & 1.21    & -10.89  & -54.56 \\
4                       & C=1.5                   & 5.87    & 4.1     & 3.02    & 0.85    & -3.87   & -41.73 \\
5                       & PM                      & -0.95   & -0.95   & -0.95   & -0.95   & -0.95   & -19.95 \\
$\ldots$(200+)                 & $\ldots$                       & $\ldots$       & $\ldots$       & $\ldots$       & $\ldots$       & $\ldots$       & $\ldots$      \\ \bottomrule

\end{tabular}
  }
    \label{tab:belief_actions}
\end{table}
Furthermore, in Table \ref{tab:belief_actions}, we list out five sample beliefs, and the generated actions obtained by the algorithm. It can be found that when the system is in better operating conditions, i.e., higher belief in states 1 and 2, the algorithm would suggest lower capacities. For the belief with larger values in intermediate states, a larger capacity is suggested. For the system approaching the failure while still in operation, the algorithm will recommend conducting PM. 

\begin{figure}
    \centering
    \includegraphics[width=0.9\linewidth]{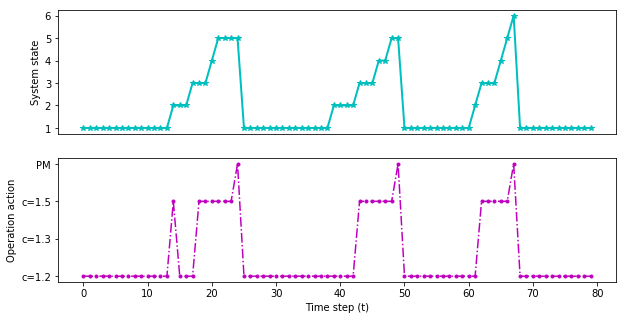}
    \caption{Illustrative examples on system states and corresponding actions}
    \label{fig:model_application}
\end{figure}

Besides, in Figure \ref{fig:model_application}, the change of the actions and system states as time elapses are presented. It can be found that when the system is in good operating states, i.e., state 1 and state 2, the system is tended to maintain lower capacity. However, when the system is approaching inferior states, i.e., state 3, 4, and 5, larger capacities are suggested. The system then is recovered to the best operating activities through maintenance when it is almost to failure. As is indicated from the transition matrices, when the system is in state 1 or 2, the probability that the system retains the same operating state is significantly higher under the operating condition $c=1.2$  is higher than the other two conditions (i.e., 0.9330 in state 1 of $\mathbf{A_1}$ comparing to 0.5606 for condition $c=1.5$ in matrix $\mathbf{A_3}$). By using the lower capacity, the degradation process is slow enough to generate more rewards than the other two operating conditions. For state $4$ and $5$, however, condition $c=1.5$ has larger transition probabilities than $c=1.2$. The algorithm jointly considers the degradation process and the instant capacity rewards, which results in the decision for adapting the condition with the largest capacity.

\begin{table}
\centering

   \caption{Sample belief values and its corresponding actions}
   \resizebox{\columnwidth}{!}{%
\begin{tabular}{@{}c|ccccccc@{}}
\toprule
\multirow{2}{*}{Sample   ID} & \multirow{2}{*}{Action} & \multicolumn{6}{c}{Belief}                               \\ \cmidrule(l){3-8} 
                             &                         & State 1 & State 2 & State 3 & State 4 & State 5 & F      \\ \midrule
1                            & C=1.2                   & 0.8185  & 0.0888  & 0.0927  & 0       & 0       & 0      \\
2                            & C=1.5                   & 0       & 0.2227  & 0.6484  & 0.0931  & 0.0146  & 0.0213 \\
3                            & C=1.5                   & 0       & 0       & 0.3551  & 0.2606  & 0.3169  & 0.0674 \\
4                            & PM                      & 0       & 0       & 0       & 0.3373  & 0.1655  & 0.4972 \\
5                            & PM                      & 0       & 0       & 0       & 0       & 0.6291  & 0.3709 \\ \bottomrule
\end{tabular}
 }
    \label{tab:belief_samples}
\end{table}

We then conduct simulation experiments to compare the performance of the POMDP policies with the other policies, i.e., simply following a single condition. 500 simulation experiments are conducted, with 50,000 cycle times in each simulation. The results are shown as  in Table \ref{tab:multi_performance}.
\begin{table}[]
    \centering
    \caption{Performance comparisons}
    \begin{tabular}{c|cccc}
\hline
         &	Proposed Method	&C1&	C2&C3\\
         \hline
Mean&	14617.09&11127.31&	5846.04&	-9213.23	\\
Std&	630.71&	653.41&616.38	&774.27	\\
\hline
    \end{tabular}
    
    \label{tab:multi_performance}
\end{table}
It can be found out that the POMDP is an effective framework on maintaining high performance of the manufacturing system with different operating capacities and maintenance options. 

To implement the model in real-time with the degradation signals, at the beginning of a cycle, by collecting and transforming the real-time degradation signals, features can be abstracted and are used as the observation input. Then by using the $\mathbf{\alpha}$-vectors, the matching belief states can be identified and the corresponding actions can be obtained. If PM is needed, then the bearing replacement is conducted. Otherwise, if the capacity switch is needed, then the operator will change the machine capacity following the obtained action. After the action, new signals will be captured and the above procedures will be repeated.


\subsection{Case Study 2: NASA C-MAPSS Turbofan Dataset}
\label{subsec:case_study_2}

To further evaluate the generality of the proposed degradation modeling framework, we conducted experiments on the NASA C\textsc{-}MAPSS FD001 turbofan dataset \cite{saxena2008turbofan}. FD001 represents a distinct application domain with different sensing modalities, operating dynamics, and degradation mechanisms compared with the previous case. The dataset contains run-to-failure trajectories of multiple engines operating under a single condition, with RUL as the predicted target.

\subsubsection{Dataset Introduction}
\label{subsec:nasa_dataset_intro}

Aircraft engines are typical examples of complex modern engineering systems. The NASA C-MAPSS (Commercial Modular Aero-Propulsion System Simulation) dataset simulates realistic aircraft engine degradation through a physics-based model that incorporates various failure modes and operational conditions. C-MAPSS is the software that simulates the degradation signals of commercial-grade turbofan aircraft engines. Each engine starts with varying degrees of initial wear and unknown manufacturing variations. The software collects 21 degradation signals and three operational variables (i.e., accounting for six working conditions) at every cycle time. Table \ref{tab:nasa_sensors} provides a detailed description of sensor signals.

The C-MAPSS dataset contains four sub-datasets: FD001, FD002, FD003, and FD004. In this work, we employ the FD001 sub-dataset for the evaluation of our proposed framework. FD001 contains 100 training units and 100 test units, with each unit operating under a single operating condition and experiencing one failure mode. The signals for the training units are collected from the beginning until failure, while those for the testing units are truncated at a random point before failure. The training units have a large number of cycle time records and thus show clear degradation trends. On the other hand, the test units have partial degradation trends due to the random truncation. Our primary goal is to identify the failure modes of both training units and test units as well as to predict the RUL of test units by using the available sensor signal data and inferred failure modes. Note that the actual failure modes for both training and test units are unknown.

FD001 represents a distinct application domain with different sensing modalities, operating dynamics, and degradation mechanisms compared with the previous case. The dataset contains run-to-failure trajectories of multiple engines operating under a single condition, with RUL as the predicted target. This dataset is particularly valuable for evaluating the proposed framework's ability to handle multi-sensor data and capture complex degradation patterns in a real-world application context.

\begin{table}[ht]
\centering
\caption{Detailed description of NASA C-MAPSS sensors}
\label{tab:nasa_sensors}
\resizebox{\columnwidth}{!}{%
\begin{tabular}{|l|l|l|}
\hline 
\textbf{Symbol} & \textbf{Description} & \textbf{Units} \\
\hline 
T2 & Total temperature at fan inlet & $^{\circ}$R \\
\hline 
T24 & Total temperature at LPC outlet & $^{\circ}$R \\
\hline 
T30 & Total temperature at HPC outlet & $^{\circ}$R \\
\hline 
T50 & Total temperature at LPT outlet & $^{\circ}$R \\
\hline 
P2 & Pressure at fan inlet & psia \\
\hline 
P15 & Total pressure in bypass-duct & psia \\
\hline 
P30 & Total pressure at HPC outlet & psia \\
\hline 
Nf & Physical fan speed & rpm \\
\hline 
Nc & Physical core speed & rpm \\
\hline 
epr & Engine pressure ratio (P50/P2) & - \\
\hline 
Ps30 & Static pressure at HPC outlet & psia \\
\hline 
phi & Ratio of fuel flow to Ps30 & pps/psi \\
\hline 
NRf & Corrected fan speed & rpm \\
\hline 
NRc & Corrected core speed & rpm \\
\hline 
BPR & Bypass ratio & - \\
\hline 
farB & Burner fuel-air ratio & - \\
\hline 
htBleed & Bleed enthalpy & - \\
\hline 
Nf\_dmd & Demanded fan speed & rpm \\
\hline 
PCNfR\_dmd & Demanded corrected fan speed & rpm \\
\hline 
W31 & HPT coolant bleed & lbm/s \\
\hline 
W32 & LPT coolant bleed & lbm/s \\
\hline
\end{tabular}
}
\end{table}

\subsubsection{Data Preprocessing and Feature Extraction}
\label{subsec:nasa_feature_extraction}

This subsection describes the data preprocessing and feature extraction methodology for the NASA C-MAPSS FD001 turbofan dataset. The process involves two main steps: (1) data preprocessing to clean and prepare the multi-sensor measurements for analysis, including sensor selection and RUL label construction, and (2) feature extraction and parameter estimation using the constrained IOHMM framework to learn the degradation dynamics from the engine sensor data. This methodology enables the transformation of raw engine sensor measurements into meaningful degradation states that can be used for RUL prediction and maintenance policy optimization.

The NASA C-MAPSS FD001 dataset includes 21 sensor measurements that capture different aspects of engine performance, including temperatures, pressures, speeds, and flow rates. These sensors provide rich information about the engine's health state, allowing for robust degradation modeling. However, several non-informative sensors provide little information about the degradation status of the system, and thus, they should be dropped before analysis. In particular, a sensor will be removed if it satisfies any of the following conditions:
\begin{itemize}
    \item a sensor only contains a single value,
    \item a sensor has 50\% or more missing values,
    \item a sensor with an extremely low standard deviation in its measurements (less than 0.01).
\end{itemize}

Next, we construct the RUL labels in the training set by subtracting the signal measurement time from the recorded failure time. Specifically, if unit 1 fails after 192 cycles, its RUL would be 191 at the first cycle, 190 at the second cycle, and so forth. The RUL progressively decreases from the initial cycle to 0 when the failure finally occurs. Note that the RUL of the test data is labeled using the same approach.

The proposed constrained IOHMM framework is then applied to extract meaningful degradation states from these multi-dimensional sensor data.

\revised{
\textbf{Parameter Estimation Results.} The learned left-to-right transition matrix for FD001 exhibits strong self-transition with small forward-transition probabilities that increase in later states, consistent with progressively accelerating degradation. This structure is consistent with a monotone, no-jump degradation process culminating in an absorbing failure state:

\[
\mathbf{A}=\begin{bmatrix}
0.9887 & 0.0113 & 0 \\
0      & 0.9722 & 0.0278 \\
0      & 0      & 1 \\
\end{bmatrix}.
\]
}

\subsubsection{Results and Analysis}
\label{subsec:nasa_results_analysis}

This subsection presents the experimental results and analysis for the NASA C-MAPSS FD001 turbofan dataset. The analysis demonstrates the framework's performance across three key dimensions: (1) RUL prediction performance, showing how the model accurately forecasts remaining useful life with appropriate uncertainty quantification; (2) state estimation and interpretation, revealing how the learned degradation states correspond to physical engine health conditions; and (3) maintenance policy evaluation, illustrating how the optimal control policy balances operational efficiency with maintenance costs. These results validate the generalizability of the proposed framework across different industrial domains and degradation mechanisms.

\revised{
\textbf{RUL Prediction Performance.} Figure~\ref{fig:fd001_rul} shows the predicted median RUL trajectory for a representative engine. The prediction closely follows the ground-truth RUL, displaying a conservative bias early in life and rapid convergence as failure approaches. The 95\% credible band remains tight through mid-life and widens moderately near end-of-life, reflecting increased uncertainty as remaining cycles shorten.

\begin{figure}[h]
  \centering
  \includegraphics[width=0.65\linewidth]{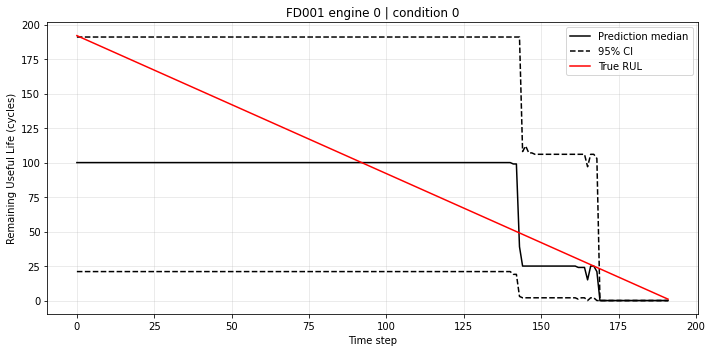}
  \caption{RUL Prediction on FD001 dataset.}
  \label{fig:fd001_rul}
\end{figure}

\textbf{State Estimation and Interpretation.} In addition, Figure \ref{fig:mean_fd001} illustrates representative raw sensor trajectories with three FD001 engines, each panel showing the time series of three sensors (black lines) together with the state-dependent emission means learned by our proposed model (colored dashed lines). Each dashed level reflects the typical magnitude of that sensor when the system occupies a particular hidden state. Because the emission model is a simple linear regression within each state, these mean levels provide an interpretable signature of health progression: lower states correspond to smaller sensor readings, while higher states exhibit progressively larger values. The observed monotonic rise of the raw signals causes the inferred states to align from "healthy/low-reading'' regimes to "high-reading/degraded'' regimes, with the highest-mean state consistently appearing near the end of life. This correspondence between increasing sensor amplitude and latent-state ordering supports the physical interpretability of the learned degradation process.

\begin{figure}[h]
  \centering
  \includegraphics[width=0.85\linewidth]{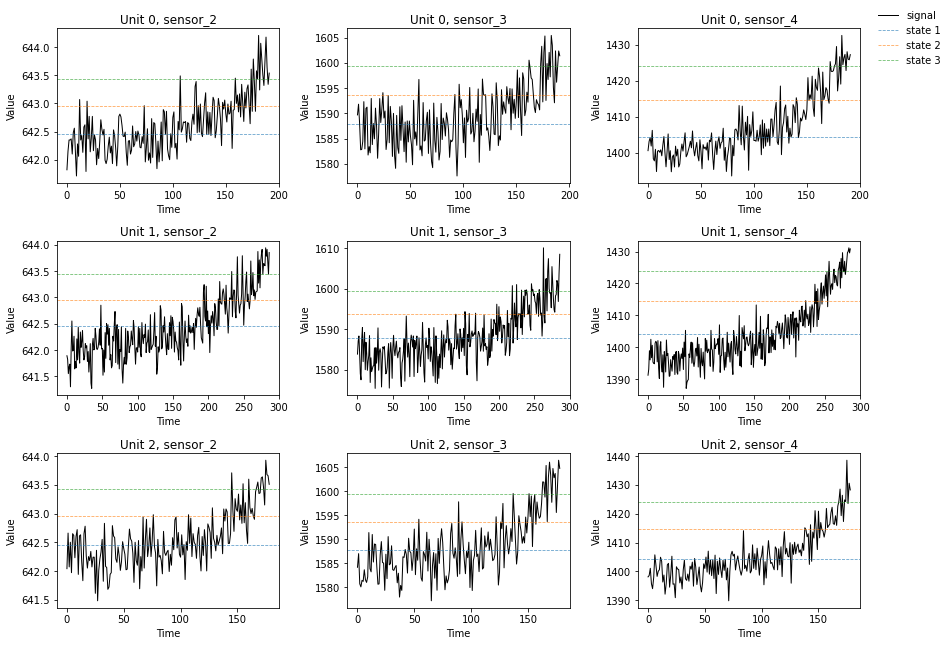}
  \caption{State estimation for FD001 dataset.}
  \label{fig:mean_fd001}
\end{figure}

\begin{figure}[h]
  \centering
  \includegraphics[width=0.85\linewidth]{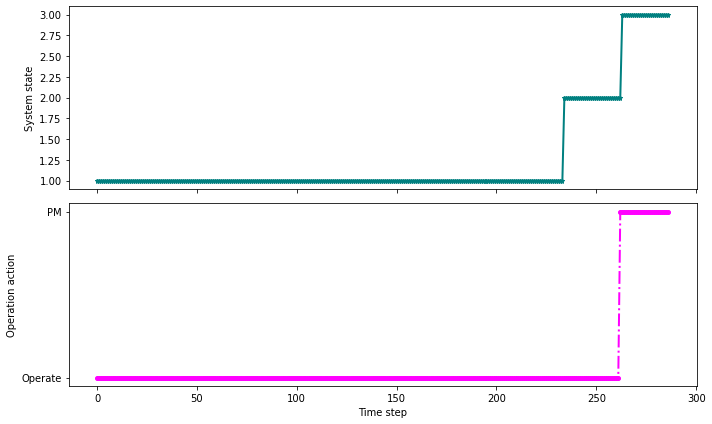}
  \caption{Illustrative examples on system states and corresponding actions.}
  \label{fig:control}
\end{figure}

Besides, in Figure \ref{fig:control}, the evolution of the system states and the corresponding maintenance actions over time is illustrated.
It can be observed that the system generally operates in the healthy states (state 1 and state 2) for extended periods, during which no action is taken and normal operation continues.
When the degradation process drives the system toward the inferior state 3, the policy triggers a preventive maintenance (PM) action, which immediately repairs the system back to the healthiest state 1.
This pattern reflects the learned trade-off: remaining in low-degradation states yields steady long-term rewards, while timely PM avoids the rapid reward loss that would occur if the system were allowed to stay in the high-risk state.

Table~\ref{tab:policy_fd001} reports the optimal actions obtained by evaluating
the POMDP value function on a discretized belief space of 200 points.
For each belief $b$ over the three latent states we compute
$Q(b,\text{Operate})$ and $Q(b,\text{PM})$ and select
$a^\star(b)=\arg\max_{a} Q(b,a)$.
The last column lists $V^{\star}(b)=\max_{a} Q(b,a)$ and the
\texttt{Action} column shows the corresponding $a^\star(b)$.

The three Value function'' columns provide the action--conditional returns that
would be obtained if the true latent state were known.
When the chosen action is \emph{Operate} (top block), the state-wise values are
approximately $4.509$ (State~1), $-0.696$ (State~2), and $-80.401$ (State~3).
When the chosen action is \emph{PM} (bottom block), the state-wise values are
about $-2.649$ (States~1--2) and $-21.648$ (State~3).
Thus, operating is attractive in healthy states (1--2) but catastrophic in the
degraded state~3, whereas PM imposes a moderate cost in healthy states and
substantially mitigates the loss in state~3 relative to operating.
Consistent with these trade-offs, the optimal policy operates when the belief
places little mass on state~3 (e.g., Belief IDs 0--4 with
$V^\star(b)\approx 4.509$) and switches to PM once the posterior probability of
state~3 becomes large (e.g., Belief IDs 195--199 with
$V^\star(b)\approx -21.642$).

\begin{table}[t]
\centering
\caption{Representative belief points and optimal actions with state--wise value functions.}
\label{tab:policy_fd001}
\resizebox{\columnwidth}{!}{%
\begin{tabular}{@{}c|c|rrrr@{}}
\toprule
Belief ID & Action &
\multicolumn{4}{c}{Value function} \\
& & \multicolumn{1}{c}{State 1}
  & \multicolumn{1}{c}{State 2}
  & \multicolumn{1}{c}{State 3}
  & \multicolumn{1}{c}{Total $V^\star(b)$} \\
\midrule
0    & Operate &  4.5091 &  -0.6955 & -80.401 &  \phantom{-}4.5090\\
1    & Operate &  4.5091 &  -0.6955 & -80.401 &  \phantom{-}4.5089 \\
2    & Operate &  4.5091 &  -0.6955 & -80.401 &  \phantom{-}4.5089 \\
3    & Operate &  4.5091 &  -0.6955 & -80.401 &  \phantom{-}4.5088 \\
4    & Operate &  4.5091 &  -0.6834 & -79.964 &  \phantom{-}4.5085 \\
\midrule
\ldots & \ldots & \ldots & \ldots & \ldots & \ldots \\
195 & PM      & -2.6486 & -2.6486 & -21.648 & -21.642 \\
196 & PM      & -2.6486 & -2.6486 & -21.648 & -21.642 \\
197 & PM      & -2.6486 & -2.6486 & -21.648 & -21.642 \\
198 & PM      & -2.6486 & -2.6486 & -21.648 & -21.642 \\
199 & PM      & -2.6486 & -2.6486 & -21.648 & -21.642 \\
\bottomrule
\end{tabular}%
}
\end{table}

Together with our XJTU-SY results, FD001 demonstrates that the proposed framework \emph{generalizes across equipment types and degradation modes}. The model recovers physically plausible state dynamics and yields calibrated RUL predictions with quantified uncertainty in an aero-engine setting, supporting the method's effectiveness and adaptability beyond bearings.
}

\section{Sensitivity Analysis}
\label{sec:sensitivity_analysis}
\label{sec:results_sensitivity_analysis}

This section presents comprehensive analysis of the proposed framework's performance, including sensitivity analysis of key hyperparameters, comparison with baseline methods, and evaluation of the robustness of the learned models. The analysis is conducted primarily on the XJTU-SY bearing dataset to provide detailed insights into the framework's behavior and performance characteristics.

In this section, we discuss the key implications of the proposed framework, including the effects of capacity-dependent degradation, the influence of critical hyperparameters (such as the number of hidden states), and the robustness of the model to parameter variations. We also highlight practical insights for maintenance decision-making and potential directions for future work.

\revised{  
\subsection{Choice of Hidden-State Number $K$}
\label{subsec:choice_hidden_states}

A critical modeling choice in the proposed constrained IOHMM is the number of hidden states $K$.
In the absence of strong prior knowledge, $K$ is typically selected using model selection criteria that balance goodness of fit against model complexity. A model with too few states may underfit the data and fail to capture meaningful degradation phases, whereas an excessively large number of states risks overfitting by introducing redundant or statistically indistinguishable states. To address this, we fitted traditional HMMs with candidate state numbers ($K=4$ to $K=8$) and compared their performance using the Akaike Information Criterion (AIC) and Bayesian Information Criterion (BIC). Both metrics penalize overly complex models while rewarding better likelihood fit. As shown in Table \ref{tab:hmm_aic_bic}, the model with $K=6$ achieves the lowest AIC (1920.04) and BIC (2368.05), indicating that six hidden states provides the best trade-off between explanatory power and parsimony. Larger models (e.g., $K=7$ or $K=8$) do not improve the likelihood sufficiently to offset the increase in the number of parameters, while smaller models ($K=4$ or $K=5$) show higher AIC/BIC values. Therefore, we selected $K=6$ as the most appropriate number of hidden states based on these information-theoretic criteria. 

\begin{table}[h]
\centering
\caption{Model selection for HMMs with different numbers of hidden states $K$.}
\label{tab:hmm_aic_bic}
\begin{tabular}{rrrrrr}
\toprule
$K$ & logL & num\_params & AIC & BIC & train\_sec \\
\midrule
4 & -1055.48 & 95  & 2300.96 & 2575.55 & 0.07 \\
5 &  -971.18 & 124 & 2190.37 & 2548.77 & 0.20 \\
\textbf{6} &  \textbf{-805.02} & \textbf{155} & \textbf{1920.04} & \textbf{2368.05} & \textbf{0.15} \\
7 &  -902.94 & 188 & 2181.87 & 2725.26 & 0.07 \\
8 &  -911.23 & 223 & 2268.45 & 2913.00 & 0.15 \\
\bottomrule
\end{tabular}
\end{table}
}

\revised{
\subsection{Impact of the Hidden-State Number $K$}
\label{subsec:impact_hidden_states}

The number of hidden states $K$ serves as an important hyperparameter in the constrained IOHMM
and directly affects both model fitting and the quality of the derived maintenance policy.
To evaluate this effect, we systematically varied $K$ from 4 to 8 and examined two key outcome
measures: the \emph{average discounted return} and the \emph{preventive maintenance (PM) action ratio}.
The results are summarized in Figure~\ref{fig:iohmm_k_sweep}.

\begin{figure}[h]
  \centering
  \includegraphics[width=0.65\linewidth]{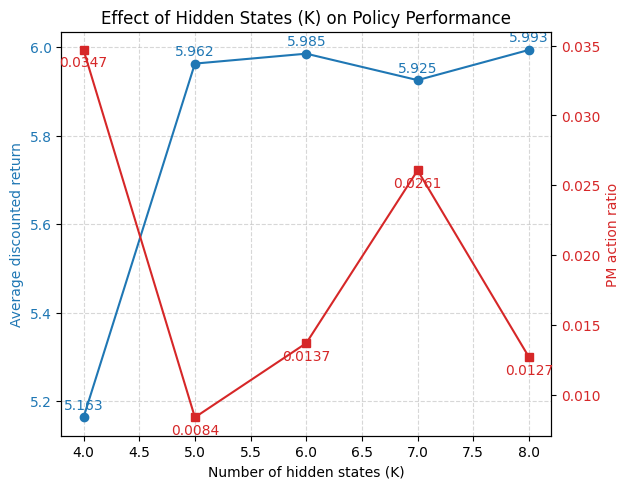}
  \caption{Effect of the number of hidden states $K$ on policy performance.
  \textbf{Blue, left axis:} average discounted return.
  \textbf{Red, right axis:} preventive maintenance (PM) action ratio (lower = fewer interventions).}
  \label{fig:iohmm_k_sweep}
\end{figure}

\begin{itemize}
    \item \textbf{Average discounted return.}  
    This metric captures the long-term reward achieved by the learned policy, reflecting the overall
    effectiveness of the maintenance strategy in balancing reliability and cost.
    Returns increase markedly from $K{=}4$ to $K{=}6$ and remain near 6.0 for $K{=}7$ and $K{=}8$,
    indicating that richer state representations enable the model to capture system dynamics
    more accurately and support more effective decision making.

    \item \textbf{Preventive maintenance (PM) action ratio.}  
    This metric measures the frequency of preventive interventions.
    Smaller models (e.g., $K{=}4$) result in higher PM ratios (0.0347), implying more frequent
    maintenance actions. At $K{=}8$, the ratio decreases to 0.0127, showing that a larger state
    space allows the policy to schedule PM more selectively while maintaining high returns.
\end{itemize}

These results demonstrate that the hidden-state number is not merely a modeling detail but
a key driver of policy behavior: smaller $K$ tends to over-prescribe maintenance with lower
overall returns, whereas larger $K$ (e.g., $K{=}7$--$8$) achieves near-optimal returns with fewer
interventions.
}

\revised{
\subsection{Sensitivity Analysis of Model Parameters}
\label{subsec:sensitivity_model_parameters}

To evaluate the robustness of the proposed framework, we conducted a dedicated sensitivity
analysis in which key model parameters---specifically the \emph{state transition probabilities} and
\emph{emission parameters}---were perturbed and re-estimated under different hidden-state settings
($K=5,6,7$). For each configuration we computed:

\begin{itemize}
    \item \textbf{Average stay/forward/backward probabilities} of the learned transition matrices,
    which indicate how strongly the model favors remaining in the current state versus progressing
    or regressing.
    \item \textbf{Average discounted return} and \textbf{preventive maintenance (PM) action ratio}
    of the derived maintenance policy, which together reflect long-term decision quality and
    intervention frequency.
\end{itemize}

\begin{table}[h]
\centering
\caption{Sensitivity analysis of model parameters and policy outcomes for different numbers of hidden states $K$.}
\label{tab:iohmm_robust}
\begin{tabular}{cccccc}
\toprule
$K$ &
Avg.\ stay &
Avg.\ forward &
Avg.\ backward &
Avg.\ discounted return &
PM action ratio \\
\midrule
5 & 0.710 & 0.089 & 0.200 & 5.845 & 0.025 \\
6 & 0.668 & 0.124 & 0.208 & 5.949 & 0.015 \\
7 & 0.661 & 0.124 & 0.214 & 5.974 & 0.022 \\
\bottomrule
\end{tabular}
\end{table}

As summarized in Table~\ref{tab:iohmm_robust}, the estimated transition dynamics remain stable
across different values of $K$: average stay probabilities are confined to a narrow range
(0.66--0.71), and forward and backward probabilities vary only slightly
(forward $\approx$0.09--0.12; backward $\approx$0.20--0.21).
Policy outcomes show similar consistency: the average discounted return remains close to
5.85--5.97, while the PM action ratio varies modestly between 0.015 and 0.025.
These results demonstrate that the maintenance recommendations are robust to moderate
changes in model parameters and to the choice of hidden-state dimension.
}

\revised{
\subsection{Comparison with Traditional HMM}
\label{subsec:comparison_traditional_hmm}

To further justify the necessity of the proposed constrained IOHMM, we conducted a direct
comparison with a conventional unconstrained (classical'') HMM that allows unrestricted state
transitions. Table~\ref{tab:hmm_condition_compare} summarizes the results across three operating
conditions.

\begin{table}[h]
\centering
\caption{Comparison of classical HMM and proposed constrained IOHMM across conditions.}
\label{tab:hmm_condition_compare}
\resizebox{0.9\textwidth}{!}{%
\begin{tabular}{ccccccc}
\toprule
\textbf{Cond.} & \textbf{Classical logL} & \textbf{Rev.\ steps} &
\textbf{Back prob.} & \textbf{Constrained logL} &
\textbf{Rev.\ steps} & \textbf{Back prob.} \\
\midrule
0 & -588.18 & 5  & 2.57 & -527.84 & 0 & 0.00 \\
1 & -641.76 & 22 & 2.77 & -549.05 & 0 & 0.00 \\
2 & -571.96 & 4  & 1.86 & -558.45 & 0 & 0.00 \\
\midrule
\textbf{Mean} & -600.63 & 10.33 & 2.40 & -545.11 & 0.0 & 0.0 \\
\bottomrule
\end{tabular}}
\end{table}

\begin{itemize}
    \item \textbf{Model fit (log likelihood).}  
    The proposed constrained IOHMM achieves competitive log-likelihood values (mean
    $-545.11$) compared with the classical HMM (mean $-600.63$), indicating that predictive
    accuracy is not sacrificed despite structural restrictions.

    \item \textbf{Reverse transitions.}  
    The classical HMM exhibits non-monotonic behavior, averaging 10.3 reverse steps per
    sequence---an outcome that is physically implausible in a degradation process.
    In contrast, the constrained IOHMM eliminates all reverse transitions, ensuring a consistent
    left-to-right progression that aligns with the known physical failure mechanism.

    \item \textbf{Backward transition probability.}  
    The classical model shows non-negligible backward transition probabilities (mean 2.40),
    whereas the constrained model enforces strict monotonic progression with zero probability
    of reversals.
\end{itemize}

These results demonstrate that although both models capture the observed data, the classical
unconstrained HMM often yields unrealistic state sequences that contradict domain knowledge.
The constrained IOHMM avoids such artifacts and provides physically interpretable degradation
trajectories without sacrificing fit quality, offering a clear advantage over simpler approaches that
do not enforce structural consistency.
}

\section{Conclusion}
\label{sec:conclusion}
In this study, we have established a comprehensive framework for determining the optimal capacity and preventive maintenance (PM) policies in scenarios where system state observations are imperfect. We constructed a semi-supervised, left-to-right constrained Input-Output Hidden Markov Model (IOHMM) to accurately estimate the parameters of deteriorating units under various operational conditions. This model leverages data from multiple sensors to accurately discern the system's state. Additionally, we innovated by integrating the traditional Partially Observable Markov Decision Process (POMDP) algorithm with the IOHMM, aiming to ascertain the most effective maintenance strategy for the system in question.

In the future, this work can be extended to more complicated dynamical systems, such as the use of Input-Output Hidden Semi-Markov Model (IOHSMM), instead of an IOHMM can be used to model more complex system dynamics. Furthermore, a continuous-state stochastic process can be used to model the degradation systems. 

\section*{Data Availability Statement}
In compliance with the Taylor \& Francis Share Upon Request data sharing policy, the data and materials supporting the findings of this study are open sourced and freely available. Specifically, the XJTU-SY bearing dataset---developed from a laboratory experimental platform (\cite{wang2018hybrid}) --- can be accessed publicly at \url{https://github.com/WangBiaoXJTU/xjtu-sy-bearing-datasets}. Additionally, the NASA C-MAPSS Dataset used in this study is publicly available online through the NASA Open Data Portal at \url{https://data.phmsociety.org/nasa/}. Should further information or additional data be required, please contact the corresponding author.

\ifblind
\else
\section*{Funding}
This work was supported by the NSF under Grant number 1922739.
\fi

\section*{Disclosure of Interest}
The authors declare that they have no known competing financial or non-financial interests that could have appeared to influence the work reported in this manuscript.

\bibliographystyle{apalike}
\bibliography{final}

@article{shen2017reliability,
  title={Reliability performance for dynamic multi-state repairable systems with K regimes},
  author={Shen, Jingyuan and Cui, Lirong},
  journal={IISE Transactions},
  volume={49},
  number={9},
  pages={911--926},
  year={2017},
  publisher={Taylor \& Francis}
}

@article{sun2020managing,
  title={Managing component degradation in series systems for balancing degradation through reallocation and maintenance},
  author={Sun, Qiuzhuang and Ye, Zhi-Sheng and Zhu, Xiaoyan},
  journal={IISE transactions},
  volume={52},
  number={7},
  pages={797--810},
  year={2020},
  publisher={Taylor \& Francis}
}

@article{kang2019performance,
  title={Performance evaluation of production systems using real-time machine degradation signals},
  author={Kang, Yunyi and Yan, Hao and Ju, Feng},
  journal={IEEE Transactions on Automation Science and Engineering},
  volume={17},
  number={1},
  pages={273--283},
  year={2019},
  publisher={IEEE}
}

@article{butler2017introduction,
  title={An introduction to predictive maintenance},
  author={Butler, J and Smalley, C},
  journal={Pharmaceutical Engineering},
  volume={37},
  number={3},
  year={2017}
}

@article{kang2019flexible,
  title={Flexible preventative maintenance for serial production lines with multi-stage degrading machines and finite buffers},
  author={Kang, Yunyi and Ju, Feng},
  journal={IISE Transactions},
  volume={51},
  number={7},
  pages={777--791},
  year={2019},
  publisher={Taylor \& Francis}
}

@article{caesarendra2011combined,
  title={Combined probability approach and indirect data-driven method for bearing degradation prognostics},
  author={Caesarendra, Wahyu and Widodo, Achmad and Thom, Pham Hong and Yang, Bo-Suk and Setiawan, Joga Dharma},
  journal={IEEE Transactions on Reliability},
  volume={60},
  number={1},
  pages={14--20},
  year={2011},
  publisher={IEEE}
}

@article{tobon2012data,
  title={A data-driven failure prognostics method based on mixture of Gaussians hidden Markov models},
  author={Tobon-Mejia, Diego Alejandro and Medjaher, Kamal and Zerhouni, Noureddine and Tripot, Gerard},
  journal={IEEE Transactions on reliability},
  volume={61},
  number={2},
  pages={491--503},
  year={2012},
  publisher={IEEE}
}

@inproceedings{kang2019joint,
  title={Joint Optimization of Operating Mode and Part Sequence for Robot Loading Process Considering Real-time Health Condition},
  author={Kang, Yunyi and Ju, Feng},
  booktitle={2019 IEEE 15th International Conference on Automation Science and Engineering (CASE)},
  pages={48--53},
  year={2019},
  organization={IEEE}
}

@article{wang2012overview,
  title={An overview of the recent advances in delay-time-based maintenance modelling},
  author={Wang, Wenbin},
  journal={Reliability Engineering \& System Safety},
  volume={106},
  pages={165--178},
  year={2012},
  publisher={Elsevier}
}

@article{wang2018hybrid,
  title={A hybrid prognostics approach for estimating remaining useful life of rolling element bearings},
  author={Wang, Biao and Lei, Yaguo and Li, Naipeng and Li, Ningbo},
  journal={IEEE Transactions on Reliability},
  volume={69},
  number={1},
  pages={401--412},
  year={2018},
  publisher={IEEE}
}

@article{saxena2008turbofan,
  title={Turbofan engine degradation simulation data set},
  author={Saxena, Abhinav and Goebel, Kai},
  journal={NASA ames prognostics data repository},
  volume={18},
  pages={878--887},
  year={2008},
  publisher={NASA Ames Research Center Moffett Field, CA, USA}
}

@article{sun2014prognostics,
  title={Prognostics uncertainty reduction by fusing on-line monitoring data based on a state-space-based degradation model},
  author={Sun, Jianzhong and Zuo, Hongfu and Wang, Wenbin and Pecht, Michael G},
  journal={Mechanical Systems and Signal Processing},
  volume={45},
  number={2},
  pages={396--407},
  year={2014},
  publisher={Elsevier}
}

@article{lu1993using,
  title={Using degradation measures to estimate a time-to-failure distribution},
  author={Lu, C Joseph and Meeker, William O},
  journal={Technometrics},
  volume={35},
  number={2},
  pages={161--174},
  year={1993},
  publisher={Taylor \& Francis}
}

@article{yan2016multiple,
  title={Multiple sensor data fusion for degradation modeling and prognostics under multiple operational conditions},
  author={Yan, Hao and Liu, Kaibo and Zhang, Xi and Shi, Jianjun},
  journal={IEEE Transactions on Reliability},
  volume={65},
  number={3},
  pages={1416--1426},
  year={2016},
  publisher={IEEE}
}

@article{rabiner2002tutorial,
  title={A tutorial on hidden Markov models and selected applications in speech recognition},
  author={Rabiner, Lawrence R},
  journal={Proceedings of the IEEE},
  volume={77},
  number={2},
  pages={257--286},
  year={2002},
  publisher={Ieee}
}

@article{soualhi2016hidden,
  title={Hidden Markov models for the prediction of impending faults},
  author={Soualhi, Abdenour and Clerc, Guy and Razik, Hubert and Guillet, Fran{\c{c}}ois and others},
  journal={IEEE Transactions on Industrial Electronics},
  volume={63},
  number={5},
  pages={3271--3281},
  year={2016},
  publisher={IEEE}
}

@article{boutros2011detection,
  title={Detection and diagnosis of bearing and cutting tool faults using hidden Markov models},
  author={Boutros, Tony and Liang, Ming},
  journal={Mechanical Systems and Signal Processing},
  volume={25},
  number={6},
  pages={2102--2124},
  year={2011},
  publisher={Elsevier}
}

@article{liu2014zero,
  title={Zero crossing and coupled hidden Markov model for a rolling bearing performance degradation assessment},
  author={Liu, Tao and Chen, Jin and Dong, Guangming},
  journal={Journal of Vibration and Control},
  volume={20},
  number={16},
  pages={2487--2500},
  year={2014},
  publisher={SAGE Publications Sage UK: London, England}
}

@article{athanasopoulou2010maximum,
  title={Maximum likelihood failure diagnosis in finite state machines under unreliable observations},
  author={Athanasopoulou, Eleftheria and Li, Lingxi and Hadjicostis, Christoforos N},
  journal={IEEE Transactions on Automatic Control},
  volume={55},
  number={3},
  pages={579--593},
  year={2010},
  publisher={IEEE}
}

@article{geramifard2013multimodal,
  title={Multimodal hidden Markov model-based approach for tool wear monitoring},
  author={Geramifard, Omid and Xu, Jian-Xin and Zhou, Jun-Hong and Li, Xiang},
  journal={IEEE Transactions on Industrial Electronics},
  volume={61},
  number={6},
  pages={2900--2911},
  year={2013},
  publisher={IEEE}
}

@article{camci2010health,
  title={Health-state estimation and prognostics in machining processes},
  author={Camci, Fatih and Chinnam, Ratna Babu},
  journal={IEEE Transactions on automation science and engineering},
  volume={7},
  number={3},
  pages={581--597},
  year={2010},
  publisher={IEEE}
}

@inproceedings{le2016competing,
  title={Competing deterioration processes modeling and RUL estimation based on factorial hidden Markov models},
  author={Le, Thanh Trung and Chatelain, Florent and B{\'e}renguer, Christophe},
  booktitle={MIMAR 2016-9th IMA International Conference on Modelling in Industrial Maintenance and Reliability},
  pages={108--113},
  year={2016},
  organization={Institute of Mathematics and its Applications}
}

@article{liu2015novel,
  title={A novel method using adaptive hidden semi-Markov model for multi-sensor monitoring equipment health prognosis},
  author={Liu, Qinming and Dong, Ming and Lv, Wenyuan and Geng, Xiuli and Li, Yupeng},
  journal={Mechanical Systems and Signal Processing},
  volume={64},
  pages={217--232},
  year={2015},
  publisher={Elsevier}
}

@article{bengio1994input,
  title={An input output HMM architecture},
  author={Bengio, Yoshua and Frasconi, Paolo},
  journal={Advances in neural information processing systems},
  volume={7},
  year={1994}
}

@article{bengio1996input,
  title={Input-output HMMs for sequence processing},
  author={Bengio, Yoshua and Frasconi, Paolo},
  journal={IEEE Transactions on Neural Networks},
  volume={7},
  number={5},
  pages={1231--1249},
  year={1996},
  publisher={IEEE}
}

@article{gonzalez2005modeling,
  title={Modeling and forecasting electricity prices with input/output hidden Markov models},
  author={Gonz{\'a}lez, Alicia Mateo and Roque, AM Son and Garc{\'\i}a-Gonz{\'a}lez, Javier},
  journal={IEEE Transactions on Power Systems},
  volume={20},
  number={1},
  pages={13--24},
  year={2005},
  publisher={IEEE}
}

@article{rocher2021iohmm,
  title={An iohmm-based framework to investigate drift in effectiveness of iot-based systems},
  author={Rocher, G{\'e}rald and Lavirotte, St{\'e}phane and Tigli, Jean-Yves and Cotte, Guillaume and Dechavanne, Franck},
  journal={Sensors},
  volume={21},
  number={2},
  pages={527},
  year={2021},
  publisher={MDPI}
}

@article{gebraeel2008prognostic,
  title={Prognostic degradation models for computing and updating residual life distributions in a time-varying environment},
  author={Gebraeel, Nagi and Pan, Jing},
  journal={IEEE Transactions on Reliability},
  volume={57},
  number={4},
  pages={539--550},
  year={2008},
  publisher={IEEE}
}

@article{bian2015degradation,
  title={Degradation modeling for real-time estimation of residual lifetimes in dynamic environments},
  author={Bian, Linkan and Gebraeel, Nagi and Kharoufeh, Jeffrey P},
  journal={Iie Transactions},
  volume={47},
  number={5},
  pages={471--486},
  year={2015},
  publisher={Taylor \& Francis}
}

@article{whitmore1997modelling,
  title={Modelling accelerated degradation data using Wiener diffusion with a time scale transformation},
  author={Whitmore, George Alex and Schenkelberg, Fred},
  journal={Lifetime data analysis},
  volume={3},
  number={1},
  pages={27--45},
  year={1997},
  publisher={Springer}
}

@article{klutke2002availability,
  title={The availability of inspected systems subject to shocks and graceful degradation},
  author={Klutke, G-A and Yang, Yoonjung},
  journal={IEEE Transactions on Reliability},
  volume={51},
  number={3},
  pages={371--374},
  year={2002},
  publisher={IEEE}
}

@article{liao2006maintenance,
  title={Maintenance of continuously monitored degrading systems},
  author={Liao, Haitao and Elsayed, Elsayed A and Chan, Ling-Yau},
  journal={European Journal of Operational Research},
  volume={175},
  number={2},
  pages={821--835},
  year={2006},
  publisher={Elsevier}
}

@article{liyanage2003towards,
  title={Towards a value-based view on operations and maintenance performance management},
  author={Liyanage, Jayantha P and Kumar, Uday},
  journal={Journal of Quality in Maintenance Engineering},
  volume={9},
  number={4},
  pages={333--350},
  year={2003},
  publisher={MCB UP Ltd}
}

@article{luo2015data,
  title={A data-driven two-stage maintenance framework for degradation prediction in semiconductor manufacturing industries},
  author={Luo, Ming and Yan, Heng-Chao and Hu, Bin and Zhou, Jun-Hong and Pang, Chee Khiang},
  journal={Computers \& Industrial Engineering},
  volume={85},
  pages={414--422},
  year={2015},
  publisher={Elsevier}
}

@article{alaswad2017review,
  title={A review on condition-based maintenance optimization models for stochastically deteriorating system},
  author={Alaswad, Suzan and Xiang, Yisha},
  journal={Reliability engineering \& system safety},
  volume={157},
  pages={54--63},
  year={2017},
  publisher={Elsevier}
}

@inproceedings{wang2018condition,
  title={Condition-based real-time production control for smart manufacturing systems},
  author={Wang, Feifan and Lu, Yan and Ju, Feng},
  booktitle={2018 IEEE 14th International Conference on Automation Science and Engineering (CASE)},
  pages={1052--1057},
  year={2018},
  organization={IEEE}
}

@article{douer1994optimal,
  title={Optimal repair and replacement in Markovian systems},
  author={Douer, Nir and Yechiali, Uri},
  journal={Stochastic Models},
  volume={10},
  number={1},
  pages={253--270},
  year={1994},
  publisher={Taylor \& Francis}
}

@article{keizer2017condition,
  title={Condition-based maintenance policies for systems with multiple dependent components: A review},
  author={Keizer, Minou CA Olde and Flapper, Simme Douwe P and Teunter, Ruud H},
  journal={European Journal of Operational Research},
  volume={261},
  number={2},
  pages={405--420},
  year={2017},
  publisher={Elsevier}
}

@article{jardine2006review,
  title={A review on machinery diagnostics and prognostics implementing condition-based maintenance},
  author={Jardine, Andrew KS and Lin, Daming and Banjevic, Dragan},
  journal={Mechanical systems and signal processing},
  volume={20},
  number={7},
  pages={1483--1510},
  year={2006},
  publisher={Elsevier}
}

@article{wang2002survey,
  title={A survey of maintenance policies of deteriorating systems},
  author={Wang, Hongzhou},
  journal={European journal of operational research},
  volume={139},
  number={3},
  pages={469--489},
  year={2002},
  publisher={Elsevier}
}

@article{yang2007maintenance,
  title={Maintenance scheduling for a manufacturing system of machines with adjustable throughput},
  author={Yang, Zimin and Djurdjanovic, Dragan and Ni, Jun},
  journal={IIE transactions},
  volume={39},
  number={12},
  pages={1111--1125},
  year={2007},
  publisher={Taylor \& Francis}
}

@article{hao2015controlling,
  title={Controlling the residual life distribution of parallel unit systems through workload adjustment},
  author={Hao, Li and Liu, Kaibo and Gebraeel, Nagi and Shi, Jianjun},
  journal={IEEE Transactions on Automation Science and Engineering},
  volume={14},
  number={2},
  pages={1042--1052},
  year={2015},
  publisher={IEEE}
}

@article{li2017study,
  title={Study of dynamic workload assignment strategies on production performance},
  author={Li, H and Parlikad, AK},
  journal={IFAC-PapersOnLine},
  volume={50},
  number={1},
  pages={13710--13715},
  year={2017},
  publisher={Elsevier}
}

@article{byon2010season,
  title={Season-dependent condition-based maintenance for a wind turbine using a partially observed Markov decision process},
  author={Byon, Eunshin and Ding, Yu},
  journal={IEEE Transactions on Power Systems},
  volume={25},
  number={4},
  pages={1823--1834},
  year={2010},
  publisher={IEEE}
}

@article{aldurgam2013optimal,
  title={Optimal joint maintenance and operation policies to maximise overall systems effectiveness},
  author={AlDurgam, Mohammad M and Duffuaa, Salih O},
  journal={International Journal of Production Research},
  volume={51},
  number={5},
  pages={1319--1330},
  year={2013},
  publisher={Taylor \& Francis}
}

@article{zhao2019semi,
  author  = {Zhao, X. and Kang, Y. and Yan, H. and Ju, F.},
  title   = {{Semi-supervised Constrained Hidden Markov Model Using Multiple Sensors for Remaining Useful Life Prediction and Optimal Predictive Maintenance: for Remaining Useful Life Prediction and Optimal Predictive Maintenance.}},
  journal = {Annual Conference of the {PHM} Society},
  year    = {2019},
  volume  = {11},
  number  = {1},
  note    = {\url{https://doi.org/10.36001/phmconf.2019.v11i1.851}},
}

@article{deep2023partially,
  title={Partially observable Markov decision process-based optimal maintenance planning with time-dependent observations},
  author={Deep, Akash and Zhou, Shiyu and Veeramani, Dharmaraj and Chen, Yong},
  journal={European journal of operational research},
  volume={311},
  number={2},
  pages={533--544},
  year={2023},
  publisher={Elsevier}
}

@phdthesis{gu2016real,
  title={Real-Time Maintenance Policies in Manufacturing Systems.},
  author={Gu, Xi},
  year={2016}
}

@book{puterman2014markov,
  title={Markov decision processes: discrete stochastic dynamic programming},
  author={Puterman, Martin L},
  year={2014},
  publisher={John Wiley \& Sons}
}

@inproceedings{pineau2005point,
  title={Point-based approximations for fast POMDP solving},
  author={Pineau, Jolle and Gordon, Geoffrey and Thrun, Sebastian},
  booktitle={Proc. int. symp. on robotics research},
  year={2005}
}

\vfill


\end{document}